\documentclass[graybox]{svmult}

\usepackage{mathptmx}       % selects Times Roman as basic font
\usepackage{helvet}         % selects Helvetica as sans-serif font
\usepackage{courier}        % selects Courier as typewriter font
\usepackage{type1cm}        % activate if the above 3 fonts are
\usepackage{makeidx}         % allows index generation
\usepackage{graphicx}        % standard LaTeX graphics tool
\usepackage{multicol}        % used for the two-column index
\usepackage[bottom]{footmisc}% places footnotes at page bottom

\usepackage{nicefrac}
\usepackage{amsmath}
\usepackage{amssymb}

\makeindex             % used for the subject index
\begin{document}

\title*{Hydrogen-rich Core Collapse Supernovae}
\author{Iair Arcavi}
\institute{Iair Arcavi \at Las Cumbres Observatory Global Telescope 6740 Cortona Dr. Suite 102, Goleta, CA, 93117, USA, \email{iarcavi@lcogt.net}
\at Kavli Institute for Theoretical Physics, University of California, Santa Barbara, CA 93106, USA, \email{arcavi@kitp.ucsb.edu}}
\maketitle

\abstract*{Hydrogen-rich core collapse supernovae, known as ``Type II'' supernovae, are the most common type of explosion realized in nature. They are defined by the presence of prominent hydrogen lines in their spectra. Type II supernovae are observed only in star-forming galaxies, and several events have been directly linked to massive star progenitors. Five main subclasses are identified: Type IIP (displaying a plateau in their light curve), Type IIL (displaying a light curve decline), Type IIn (displaying narrow emission lines), Type IIb (displaying increasingly strong He features with time) and 87A-likes (displaying long-rising light curves similar to that of SN\,1987A). Type IIP supernovae have been robustly established as the explosions of red supergiants, while the progenitors of Type IIL's remain elusive. Type IIn's are likely linked to luminous blue variables, Type IIb progenitors may be interacting binary systems and the prototype of the 87A-like class was observed to be the explosion of a blue supergiant. The diversity in progenitor mass, metallicity, binarity and rotation is likely responsible for the diversity in observed explosion types, but the connection between progenitor parameters and supernova properties is not yet entirely understood theoretically nor fully mapped observationally. New observational methods for constraining this connection are currently being implemented, including the analyses of large samples of events, making use of very early data (obtained hours to days from explosion) and statistical studies of host-galaxy properties.}

\abstract{Hydrogen-rich core collapse supernovae, known as ``Type II'' supernovae, are the most common type of explosion realized in nature. They are defined by the presence of prominent hydrogen lines in their spectra. Type II supernovae are observed only in star-forming galaxies, and several events have been directly linked to massive star progenitors. Five main subclasses are identified: Type IIP (displaying a plateau in their light curve), Type IIL (displaying a light curve decline), Type IIn (displaying narrow emission lines), Type IIb (displaying increasingly strong He features with time) and 87A-likes (displaying long-rising light curves similar to that of SN\,1987A). Type IIP supernovae have been robustly established as the explosions of red supergiants, while the progenitors of Type IIL's remain elusive. Type IIn's are likely linked to luminous blue variables, Type IIb progenitors may be interacting binary systems and the prototype of the 87A-like class was observed to be the explosion of a blue supergiant. The diversity in progenitor mass, metallicity, binarity and rotation is likely responsible for the diversity in observed explosion types, but the connection between progenitor parameters and supernova properties is not yet entirely understood theoretically nor fully mapped observationally. New observational methods for constraining this connection are currently being implemented, including the analyses of large samples of events, making use of very early data (obtained hours to days from explosion) and statistical studies of host-galaxy properties.}

\section{Introduction}
\label{sec:intro}

The first identification of supernovae (SNe) with hydrogen features in their spectra was reported by Minkowski (1941) \cite{Minkowski1941} who also suggested the ``provisionary'' name of ``Type II'' SNe (distinct from ``Type I'' SNe which do not display hydrogen in their spectra). To this day, we still refer to H-rich SNe as ``Type II''. They are defined observationally by the presence of broad (few thousand km\,s$^{-1}$) hydrogen lines (most notable of which is the Balmer H$\alpha$ line at rest wavelength $6563\textrm{\AA}$) in their spectra.

Type II SNe are exclusively found in star-forming galaxies, and several direct pre-explosion progenitor images reveal supergiant stars at the positions of Type II SNe  (see ``Supernova Progenitors Observed with HST''). SNe II are thus considered to be explosions of massive ($\gtrsim8M_{\odot}$) stars which have retained some or all of their hydrogen envelope prior to explosion.

\section{Subclasses}
\label{sec:subclasses}

Five main subclasses of Type II SNe have been identified. Initially, Type II SNe were divided into {\bf IIP} events (for ``plateau'') displaying constant luminosity for approximately 100 days in their light curves, and {\bf IIL} events (for ``linear'') displaying a linear decline in their light curves (in magnitude space) \cite{Barbon1979}. This nomenclature persists until today, though it is still being debated whether the explosions span a continuous class of light curve decline rates or whether IIP and IIL are indeed two populations. These two classes are often together referred to as ``normal Type II SNe'' or just ``Type II SNe''. Here I will use the term ``Type II SN'' to refer to all H-rich subclasses (not just IIP and IIL SNe).

The third subclass, Type {\bf IIn}, is identified by narrow (few hundred km\,s$^{-1}$) hydrogen emission lines with broad bases seen in the spectra \cite{Schlegel1990}. The narrow component of the lines is attributed to slowly-moving circumstellar material ejected by the SN progenitor before explosion \cite{Chugai1991}. Type IIn SNe are themselves very diverse photometrically and are discussed in detail in  ``Interacting Supernovae''.

Some SNe have been observed to display prominent broad hydrogen lines early in their evolution but later these lines weaken and the spectra become helium-dominated. At these later times these SNe look like Type Ib events (i.e. SNe with no hydrogen but with strong signatures of helium; see ``H-poor Core Collapse Supernovae''). These intermediate II - Ib events have been classified as Type {\bf IIb} SNe. Their progenitors may have experienced an intermediate stripping level between those of H-rich SNe II and those of H-poor SNe Ib.

Finally, the closest SN in modern times, SN 1987A (see reviews \cite{Arnett1989,McCray1993}), which exploded in the Large Magellanic Cloud (LMC), was a Type II SN, but it does not belong to any of the above subclasses. It displayed broad hydrogen emission like a SN IIP, but its light curve showed a long ($\sim80$\,day) rise rather than a plateau. For lack of a better name, events similar to SN 1987A are called {\bf 87A-likes}. Such events are rare ($\sim1-5\%$ of all SNe; \cite{Smartt2009, Kleiser2011}).

The five Type II SN subclasses and their definitions are summarized in Table \ref{tab:subclasses_initial}, together with well-observed SNe considered the ``prototypical example'' of each class. SN\,1979C has been long considered the prototype of IIL SNe, however it turns out that SN\,1979C is actually different than what today is referred to as IIL's (SN\,1979C is more luminous than recent IIL's, displayed early signs of possible interaction, and may have a different light curve decline as explained in Section \ref{subsec:light_curves_iipiil}). Since the definition of a IIL is still not quite clear, there is no single SN which is considered the true representative of this class. There is also no clear prototypical event for the diverse Type IIn class (see ``Interacting Supernovae''). 

\begin{table}
\caption{Division of H-rich SNe into subclasses (all spectroscopic properties are in addition to strong hydrogen lines, which define the entire Type II SN class). Several pieces of the definitions were never formulated, making it unclear whether this subdivision is complete or exhaustive. Events considered prototypical for each class are listed, though recently it has been realized that SN\,1979C is not a typical member of the IIL subclass.}
\label{tab:subclasses_initial}       % Give a unique label
\begin{tabular}{p{1.3cm}p{4.25cm}p{4.25cm}p{1.5cm}}
\hline\noalign{\smallskip}
Subclass & Photometric Properties & Spectroscopic Properties & Prototypical Example \\
\noalign{\smallskip}\svhline\noalign{\smallskip}
IIP & Plateau in light curve & & SN\,1999em \\
IIL & Linear decline in light curve & & SN\,1979C? \\
IIn & & Narrow hydrogen lines & \\
IIb &  & H-dominated then He-dominated & SN\,1993J\\
87A-like & Long light curve rise & & SN\,1987A\\
\noalign{\smallskip}\hline\noalign{\smallskip}
\end{tabular}
\end{table}

The subdivision into IIP, IIL, IIn, IIb and 87A-like is based on a mix of spectroscopic and photometric features, and is not necessarily mutually exclusive nor complete. In addition, the defining criteria are rather vague and non-quantitative. 

Natural questions which arise are: Do IIP's and IIL's have different spectroscopic properties? What are the light curve properties of IIn's and IIb's? What defines a plateau vs. a decline of the light curve? How different are all of these classes and what are the distributions of the observed parameters? 

It is only in recent years that we are starting to answer these questions. Of course, the main motivation for quantifying the observed subclasses is to answer the bigger question: What are the different progenitor systems that lead to the different explosions and why?

\section{Progenitors}
\label{sec:progenitors}

One way to answer the progenitor question directly is to identify the pre-explosion stars in high-resolution archival imaging of SN sites. This method and what we have learned from it are discussed in detail in ``Supernova Progenitors Observed with HST'' (see also \cite{Smartt2009} and \cite{Leonard2010} for reviews). Here we summarize the main results concerning Type II SNe (see also Table \ref{tab:progenitors}). We begin with the most strongly constrained scenario and move to the least constrained one.

\subsection{IIP SNe From Single Red Supergiants}

Several progenitors of Type IIP SNe have been directly identified as single red supergiants (RSGs) with masses in the range $10-17M_{\odot}$. RSGs are stars with extended hydrogen envelopes. Such envelopes are directly responsible for the plateau seen in the light curve of the ensuing SNe (see Section \ref{sec:light_curves} and ``Light Curves of Type II Supernovae''). 

\subsection{87A-Like SNe From Blue Supergiants}

SN\,1987A, on the other hand, was the explosion of a more compact blue supergiant (BSG; see ``Progenitor of SN 1987A''). Standard stellar evolution theory did not predict stars would explode in their BSG phase. Models later showed that low metallicity, fast rotation or binarity could in fact allow stars to explode as BSGs \cite{Podsiadlowski1992}. SN\,1987A was indeed in a low metallicity galaxy (the LMC), and the complex geometry of its remnant suggests its progenitor may have had a binary companion. It may have also been the rapidly rotating product of a binary merger \cite{Hillebrandt1989, Podsiadlowski1990}. Other 87A-like SNe are also found in low-metallicity environments (see Section \ref{sec:hosts}). The small radius of a BSG compared to a RSG could explain why 87A-like SNe don't show a plateau in their light curve (see Section \ref{sec:light_curves} and ``Light Curves of Type II Supernovae'').
 
\subsection{IIb SNe From Yellow Supergiants With Binary Companions} 
 
A few Type IIb SNe progenitors have been identified as yellow supergiants (YSGs), with a binary companion seen in the case of SN\,1993J (\cite{Podsiadlowski1993} and possibly also for SN\,2008ax \cite{Crockett2008, Folatelli2015} and SN\,2011dh \cite{Folatelli2014} (but see also \cite{Maund2015}). An interacting binary as the progenitor system for Type IIb SNe may explain the low hydrogen content seen in the SN spectra resulting from envelope stripping by the companion prior to explosion. It can also explain the peculiar density structure inferred for some Type IIb progenitors from double peaked light curves (see Section \ref{subsec:light_curves_iib}). Chevalier and Soderberg (2010) \cite{Chevalier2010} suggest that both compact and extended progenitors can produce IIb SNe, producing two subtypes which they name cIIb's and eIIb's respectively.

\subsection{IIn SNe From LBVs}

One Type IIn SN progenitor was detected and identified as a luminous blue variable (LBV) \cite{GalYam2007, Gal-Yam2009}. LBV's are evolved massive stars undergoing periods of enhanced mass loss. They are most commonly considered to be in transition to a Wolf-Rayet phase. Recently, Smith and Tombleson (2014) \cite{Smith2014} suggested that LBVs are actually mass gainers in interacting massive-star binary systems, but Humphreys, Weis, Davidson and Gordon (2016) \cite{Humphreys2016} challenge this interpretation. Either way, LBV's, being very massive, experience substantial mass-loss, possibly generating a H-rich wind around them. This would explain the narrow spectral features and the long-lived and occasionally very luminous light curves of Type IIn SNe.

\subsection{IIL SNe}

No clear progenitor detections are available for IIL SNe. Elias-Rosa et al. (2010) \cite{Elias-Rosa2010} identify a possible YSG progenitor for SN\,2009kr, but that source may have actually been a compact cluster \cite{Maund2015}, and the SN may have actually been a Type IIP event \cite{Fraser2010}. Progenitor non-detection limits rule out a RSG more massive than $18\textrm{M}_{\odot}$ for SN\,1980K \cite{Thompson1982}. Elias-Rosa et al. (2011) {\cite{Elias-Rosa2011} investigate the progenitor for the IIL SN\,2009hd, but due to heavy extinction to the SN location can not determine if the progenitor was a RSG or YSG. 

\begin{table}
\caption{Likely progenitors of the different H-rich core collapse SN subclasses and the evidence for them.}
\label{tab:progenitors}       % Give a unique label
\begin{tabular}{p{1.3cm}p{1.7cm}p{4cm}p{4.3cm}}
\hline\noalign{\smallskip}
Subclass & Progenitor & Direct Evidence & Indirect Evidence \\
\noalign{\smallskip}\svhline\noalign{\smallskip}
IIP & RSG & Multiple progenitor detections & Light curve plateau indicative of a thick H envelope \\
IIL & ? & \\
IIn & LBV & Single progenitor detection$^a$ & Light curve and spectral features indicative of CSM interaction \\
IIb & \begin{tabular}[t]{@{}l@{}}YSG\\(in a binary)\end{tabular} & Few progenitor detections & Light curve and spectral features indicative of a H-deficient envelope \\
87A-like & BSG & Single progenitor detection$^b$ & Light curve shape indicative of a compact progenitor \\
\noalign{\smallskip}\hline\noalign{\smallskip}
\end{tabular}
$^a$ SN\,2005gl\\
$^b$ SN\,1987A

\end{table}

\section{Relative Rates}
\label{sec:rates}

The most recent SN rate measurements were conducted by Li et al. (2011) \cite{Li2011} using the Lick Observatory Supernova Search (see also \cite{Cappellaro1999, vandenBergh2005, Li2007, Prieto2008, Smartt2009}). While Type Ia's are the most commonly {\it observed} SN (Fig. \ref{fig:observed_rates}) due to their high luminosity (see ``Type Ia Supernovae''), Type II SNe are {\it intrinsically} the most common SN type, comprising $57\%$ of all SNe (Fig. \ref{fig:intrinsic_rates}). Of all Type II SNe, IIP's are the most common, though, as mentioned, the distinction between IIP and IIL is not yet entirely clear.

\begin{figure}
\sidecaption
\includegraphics[width=0.3\textwidth,trim={0 0 2700 150},clip]{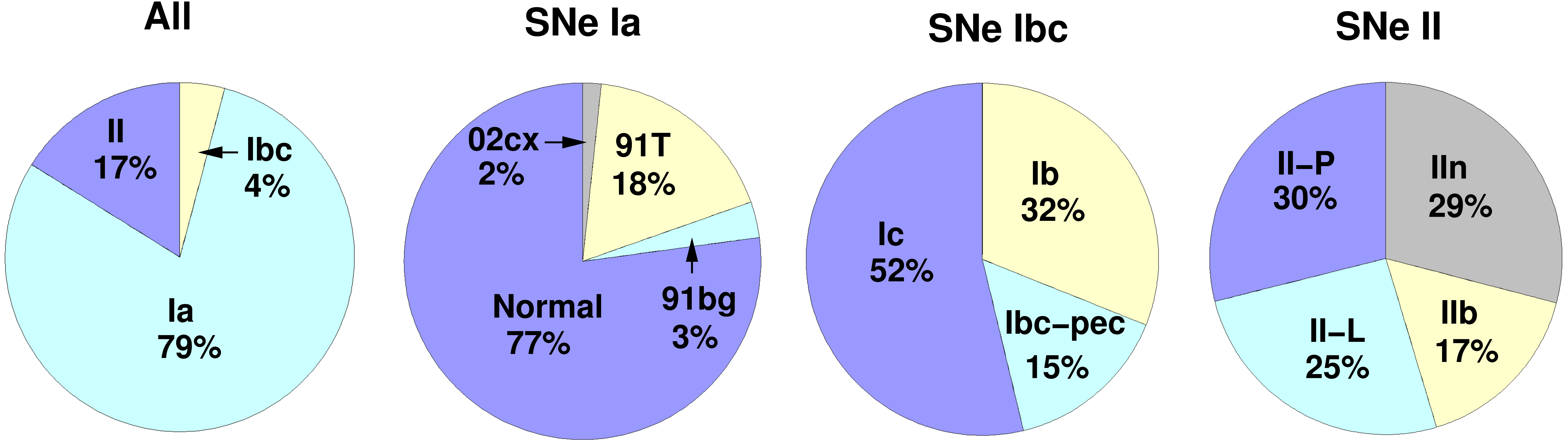}\,\,\,\,\includegraphics[width=0.3\textwidth,trim={2690 0 0 150},clip]{fig1_li_et_al_observed_rates-eps-converted-to.pdf}
\caption{{\it Observed} rates of Type II SNe compared to other SN types (left) and of the subclasses of Type II SNe (right) relative to each other, from \cite{Li2011}. Type Ia's are the brightest class of explosions considered in this chart, and are thus the most frequently observed. Figure adapted from \cite{Li2011}}
\label{fig:observed_rates}
\end{figure}

\begin{figure}
\sidecaption
\includegraphics[width=0.3\textwidth,trim={0 0 2700 150},clip]{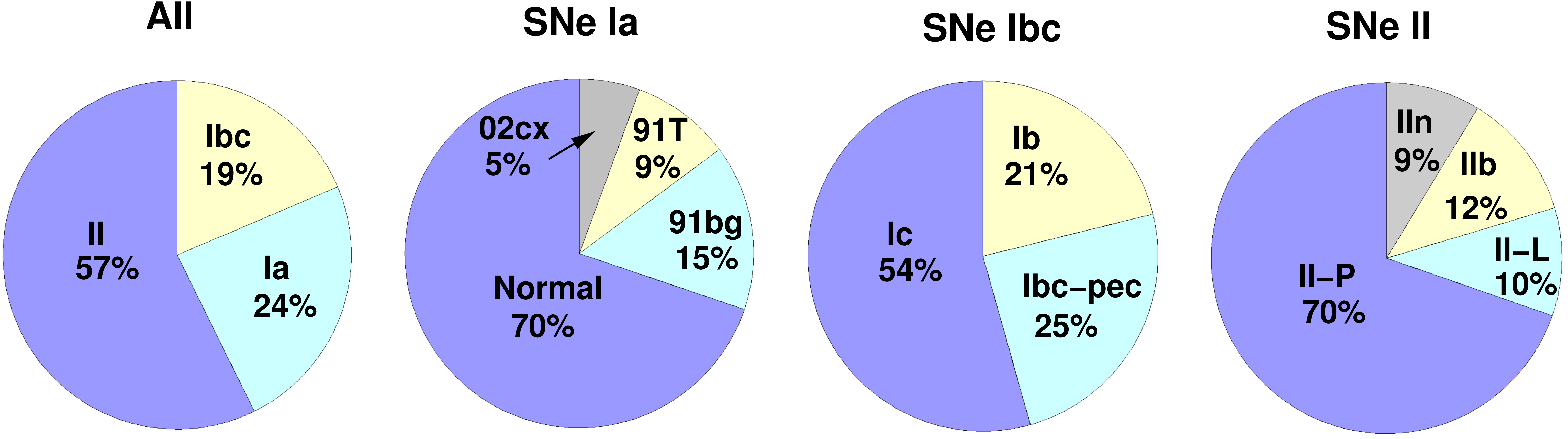}\includegraphics[width=0.3\textwidth,trim={2700 0 0 150},clip]{fig2_li_et_al_intrinsic_rates-eps-converted-to.pdf}
\caption{{\it Intrinsic} rates of Type II SNe compared to other SN types (left) and of the subclasses of Type II SNe (right) relative to each other, from \cite{Li2011}. Type II SNe are the most common type of stellar explosion, and among them, IIP's are the most common subclass (87A-likes constitute $\lesssim1-3\%$ of Type II SNe and are not considered in this plot). Figure adapted from \cite{Li2011}}
\label{fig:intrinsic_rates}
\end{figure}

Measuring intrinsic SN rates (or even relative rates) is a tricky endeavor. The question of whether a SN survey is complete out to a certain distance (i.e. that it will not miss any SNe below the survey limiting magnitude) is difficult to answer since we don't know the full luminosity distribution of all SN types. In addition, some SN types may be more affected by dust compared to others, making them more difficult to detect in the optical (see e.g. \cite{Meikle2007, Kotak2009, Meikle2011} for infrared studies of dust production in H-rich SNe). Another issue is whether different SN types prefer different galaxy types. Most SN surveys used for rates measurements were targeted surveys (in that they target known bright galaxies as potential SN hosts). Recently, untargeted surveys (which blindly search the sky) have been finding SNe in previously uncatalogued dwarf galaxies at different relative rates as in giant galaxies (see Section \ref{sec:hosts}). 

\section{Light Curves}
\label{sec:light_curves}

SN light curves encode information about the progenitor star, the explosion mechanism, the circumstellar material ejected by the progenitor prior to explosion and occasionally about the central remnant produced by the collapse of the core. Figure \ref{fig:peak_hists} displays typical $R$-band peak magnitude distributions for the different H-rich SN subtypes.

\begin{figure}
\sidecaption
\includegraphics[width=0.6\textwidth,trim={0 0 0 0},clip]{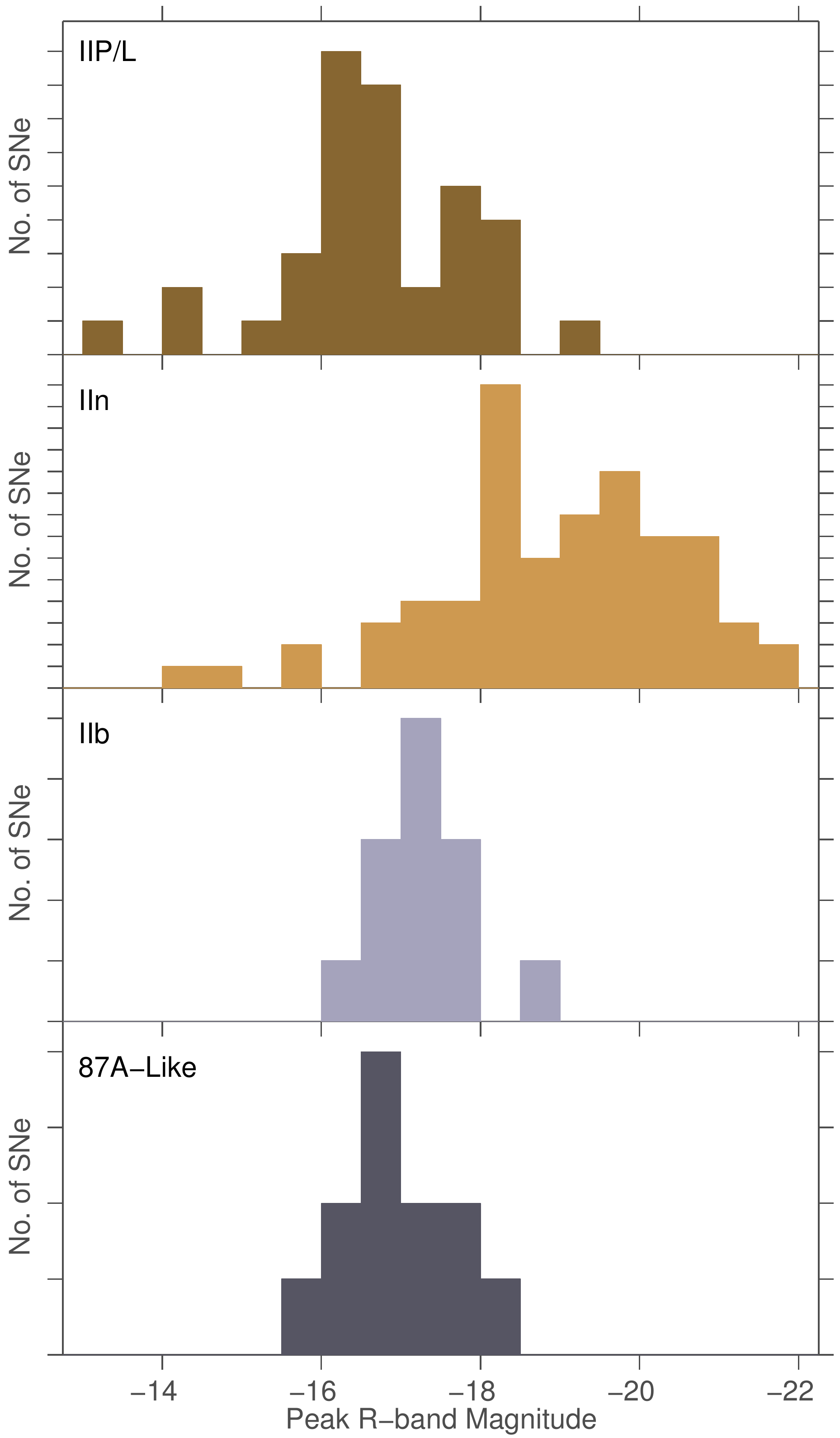}
\caption{$R$-band peak magnitudes of H-rich SNe. Type IIn events, which are powered by substantial CSM interaction, display the largest spread in peak luminosities and are the most luminous. Data from \cite{Faran2014} and Faran, Private Communication for IIP and IIL's; From \cite{Kiewe2012} and references therein, and \cite{Silverman2013} for Type IIn's; From \cite{Strotjohann2015} and references therein, and \cite{Richmond1994} for Type IIb's; From \cite{Taddia2016} and references therein for 87A-likes.}
\label{fig:peak_hists}
\end{figure}

A quantitative analysis of H-rich SN emission parameters can be found in ``Light Curves of Type II Supernovae''. Here I provide a qualitative summary of the different mechanisms and stages that can contribute to the light curve of H-rich SNe.

\begin{enumerate}
\item {\bf Shock Breakout} - As the shock (created by the rebound of the in-falling material on the newly formed proto-neutron star) reaches the surface of the star, it releases a brief (minutes to hour) pulse of X-ray and UV radiation (e.g. \cite{Falk1977,Klein1978,Waxman2007}; see also ``Shock Breakout Theory''). Shock breakout emission, generally, has been observed in only very few cases \cite{Gezari2008, Soderberg2008, Garnavich2016}.
\item {\bf Shock Cooling \& Ejecta Recombination} - The heated and ionized ejecta emit radiation as they cool and recombine (hereafter I refer to both these stages as ``cooling''). Depending on the radius, mass and density profile of the recombining hydrogen layer, this can result in a plateau of the light curve, which is the defining property of Type IIP SNe, or in a decline of the early light curve, as observed for some Type IIb SNe. Both of these cases are discussed in more detail below.
\item {\bf Radioactive Decay} - The shock deposits its energy not only in heating and ionizing the envelope, it also synthesizes heavy nuclei, some of which decay and radiate \cite{Arnett1980, Arnett1982}). The main radioactive product is $^{56}$Ni which decays to $^{56}$Co with a half-life of $\sim6$ days. $^{56}$Co then decays to the stable $^{56}$Fe with a half-life of $\sim77$ days. These processes occur through electron capture and $\beta^+$-decay, producing high energy photons (in the $\gamma$ regime). Some of these $\gamma$ photons will be down-scattered to optical wavelengths before emerging from the ejecta. The effect on the light curve depends on the amount of $^{56}$Ni produced and on the fraction of photons that are down-scattered by the ejecta (sometimes referred to as the ``degree of trapping''). The luminosity produced from radioactive decay can briefly extend the plateau of type IIP SNe \cite{Kasen2009}, is responsible for the main light curve peak in Type IIb SNe and powers the late-time light curve of all SN types. Additional radioactive elements such as $^{44}$Ti can also contribute to the emission, but at a much lower level or at much later times than the $^{56}\textrm{Ni}\rightarrow^{56}\textrm{Co}\rightarrow^{56}\textrm{Fe}$ channel.
\item {\bf CSM Interaction} - For stars exploding inside a dense circumstellar medium (CSM), additional emission is produced when the SN ejecta collide with the CSM. Such collisions tap into the vast kinetic energy of the ejecta, converting some of it to radiation. Interaction can be the main power source for Type IIn SN light curves in the optical, keeping them bright for extended periods \cite{Chevalier1977} (see ``Interacting Supernovae''). CSM interaction can also produce radio and X-ray emission.
\item {\bf Magnetar Rotational Energy} - It has been suggested \cite{Kasen2010, Woosley2010}) that rapidly rotating magnetars (with periods of less than $30$\,ms) created in core collapse events could impact the light curve. However, it is not currently known which H-rich SNe could produce such rapidly rotating magnetars, if at all.
\item {\bf Fallback Accretion} - If a black hole is formed from the collapsing core, additional material falling back on it may produce accretion luminosity \cite{Woosley1995, Zampieri1998, Dexter2013}.
\end{enumerate}

Below I focus on the roles of shock cooling (including recombination) and radioactive nickel decay in generating the different Type II SN light curves. CSM interaction is covered in ``Interacting Supernovae'', fall back accretion are discussed in ``Unusual Supernovae and Alternative Power Sources''. Magnetar power is also discussed in ``Superluminous Supernovae'', as it has been suggested as a likely power source for those extreme events. The shock breakout phase is a powerful diagnostic of SN progenitor properties, but currently very few observations of this brief phase exist. Shock breakout is discussed further in ``Shock Breakout Theory''.

\subsection{IIP/IIL SNe}
\label{subsec:light_curves_iipiil}

\subsubsection{The Plateau}

As mentioned in Section \ref{sec:progenitors}, the progenitors of SNe IIP are RSGs. RSGs are massive stars with very thick hydrogen envelopes. Here I present a simplified picture of how, after explosion, the energy deposited by the shock in the hydrogen envelope creates a plateau in the light curve. A more quantitative description was first provided by Chevalier (1976) \cite{Chevalier1976} and is discussed in ``Light Curves of Type II Supernovae''.
\begin{enumerate}
\item After core collapse, a shock wave is sent into the hydrogen envelope, ejecting, heating and ionizing the material in the envelope.
\item The ejecta expand and gradually cools from the outside inwards, forming a temperature gradient. The location inside the envelope where the temperature equals the recombination temperature of hydrogen is called the recombination front. Outside of the recombination front, the material is cooler than the recombination temperature, while inside the recombination front, the material is still ionized. Since neutral hydrogen is transparent, while ionized hydrogen is highly opaque (due to Thompson scattering off the free electrons), the recombination front is also the photosphere (i.e. photons can travel more or less freely from the recombination front outwards).
\item The first optical radiation is thus emitted from the outer ejecta without having to diffuse through a large amount of mass. This causes the light curve to rise rapidly (Fig. \ref{fig:iip_rise}). This rapid rise is not seen in stripped-envelope SNe with nickel-decay powered light curves, since that emission has to diffuse out from the center of the ejecta, where most of the nickel is created. 
\item The recombination front recedes inwards in mass through the ejecta, as the envelope expands in radius (Fig. \ref{fig:iip_diagram}). This causes the radius of the recombination front / photosphere to be roughly constant. Since the temperature at the photosphere is constant (approximately following the recombination temperature) and the radius of the photosphere is constant (due to the counter effects of it receding inside an expanding envelope), then the luminosity is roughly constant, producing a plateau in the light curve.
\item As the recombination front reaches the base of the envelope, the luminosity suddenly drops. All of the energy from the shock that was stored in the envelope has now been released. Additional photons from nickel decay, trapped in the opaque ejecta, are also instantaneously released at this stage, causing a drop in the Ni-power component as well (which is why Ni-powered 87A-like events show a drop at a approximately the same time; Fig. \ref{fig:bols}). 
\end{enumerate}
As the photosphere recedes, more and more Fe-group elements outside the photosphere re-absorb some of the light emitted at the photosphere. Fe-group elements absorb mostly blue light, thus a decline in luminosity is observed in bluer wavelengths, whereas a plateau is observed in redder wavelengths (less affected by Fe-group absorption).

\begin{figure}
\sidecaption
\includegraphics[width=\textwidth,trim={0 0 0 550},clip]{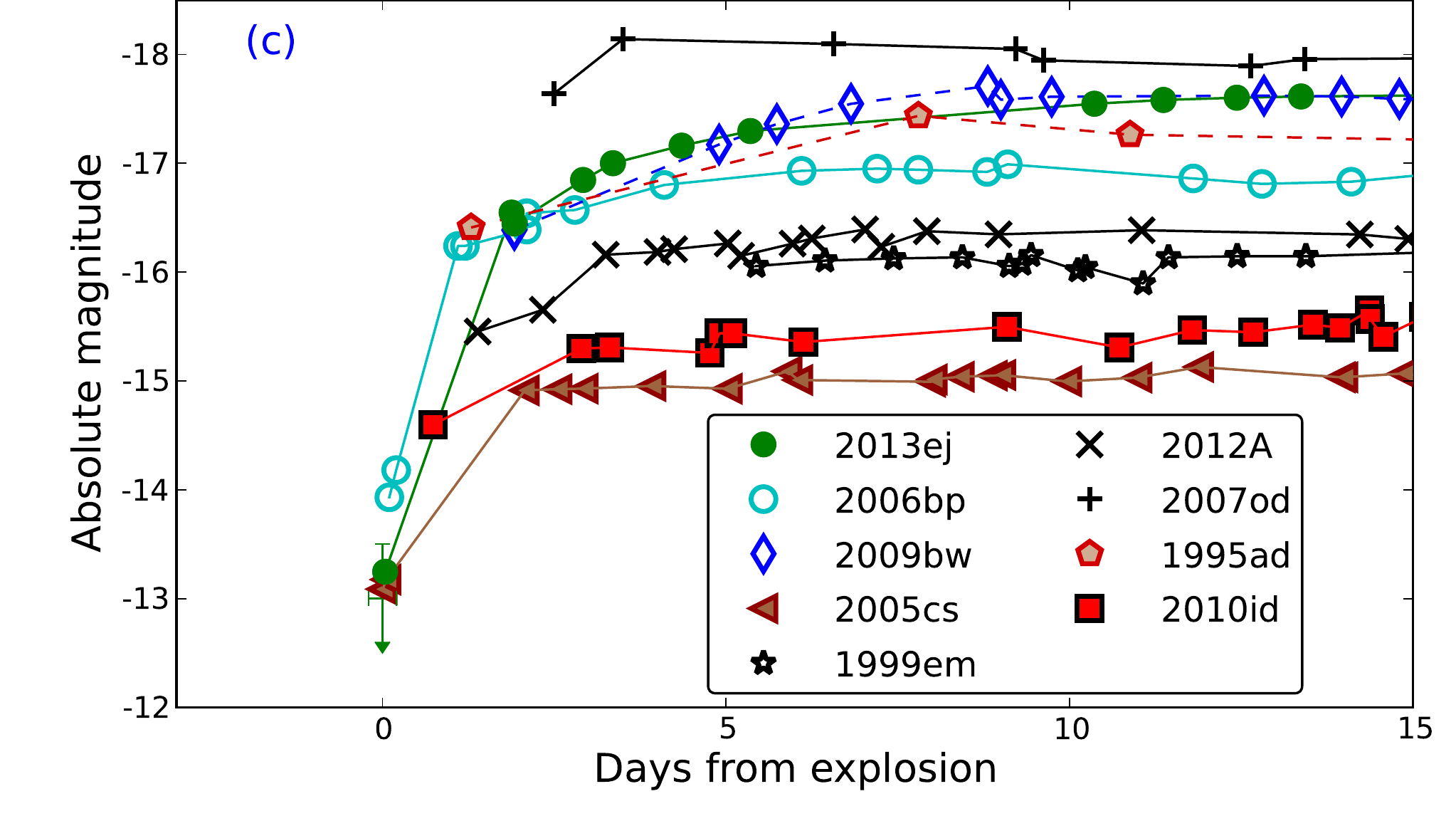}
\caption{$R$-band light curves of the rise to the plateau for Type IIP SNe from \cite{Valenti2013} and references therein. Most SNe IIP rise very quickly (within a few days) to their plateau.}
\label{fig:iip_rise}
\end{figure}

\begin{figure}
\sidecaption
\includegraphics[width=\textwidth,trim={70 100 0 100},clip]{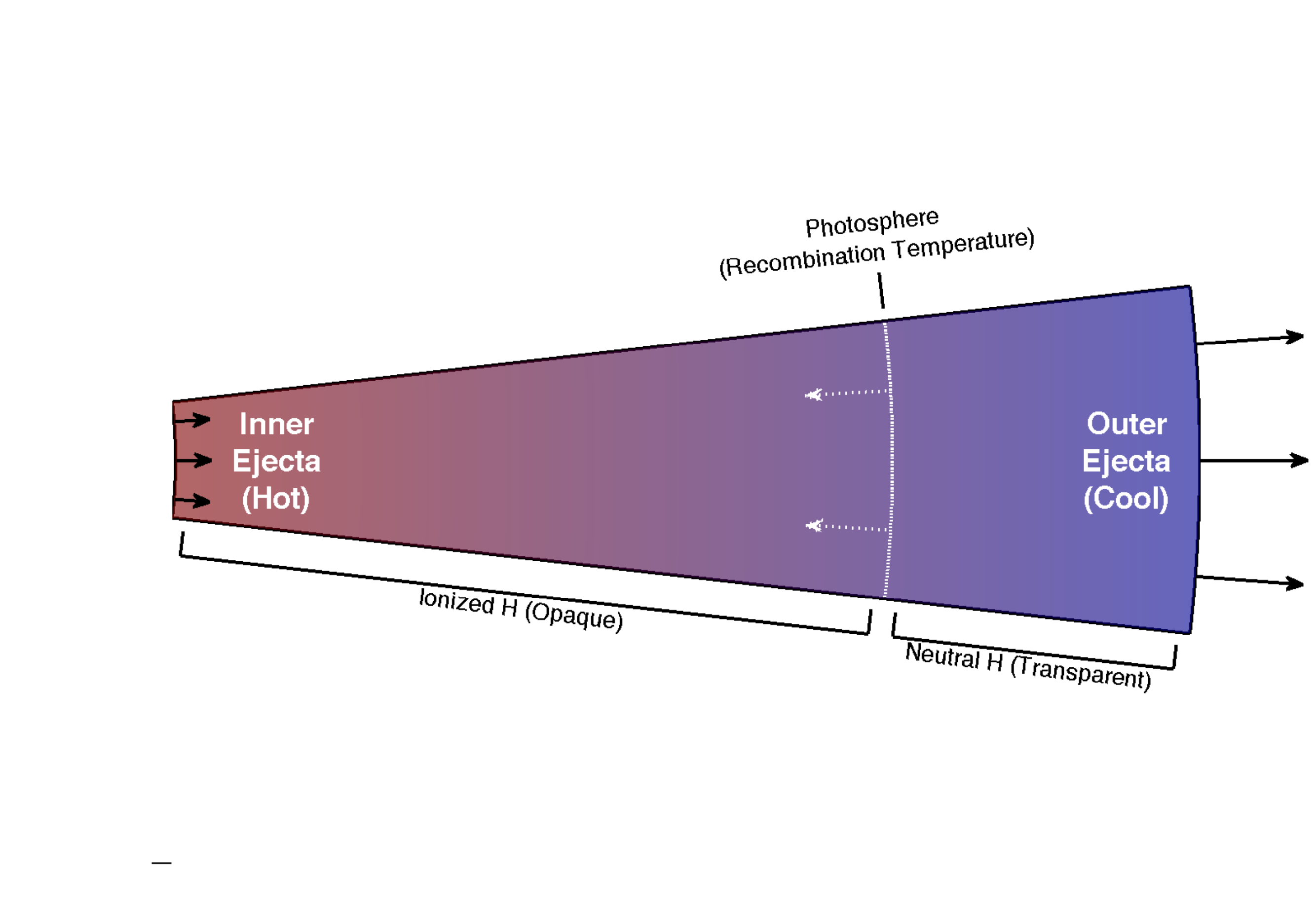}
\caption{Simplified illustration of a slab of expanding ejecta in a Type IIP SN (the direction of expansion is shown by black arrows). The outer, less dense, ejecta cools first, forming a temperature gradient. Since ionized hydrogen is opaque (due to Thompson scattering off the free electrons), the ejecta is divided into a transparent region (cooler than the recombination temperature) and an opaque region (hotter than the recombination temperature). The recombination front is therefore also approximately the photosphere (white dotted line). This front recedes in the expanding ejecta (white dotted arrows) as it cools from the outside in, maintaining a roughly constant physical radius. The constant radius and temperature that set the photosphere produce the constant luminosity that defines Type IIP SNe, until the ejecta fully recombines and all of the shock-deposited energy is released.}
\label{fig:iip_diagram}
\end{figure}

Thus, the plateau in the light curve is powered from the cooling of the ejecta which was heated by the shock. The ionized envelope acts as a reservoir of energy deposited by the shock, slowly releasing it as radiation. For reasons that will become clear later, it is not a good idea to call this phase in the light curve ``the plateau phase'', but rather we will refer to it as ``the optically thick phase'', as suggested by Anderson et al. (2014) \cite{Anderson2014}.

\subsubsection{The Nickel Decay Tail}

The amount of $^{56}$Ni produced in a Type IIP SN is $10^{-4}-10^{-2}M_{\odot}$ (e.g. \cite{Hamuy2003}, \cite{Spiro2014} and references therein). Nakar et al. (2016) \cite{Nakar2016} propose that as much as $20\%$ of the flux released during the optically thick phase is from $^{56}$Ni decay. Kasen and Woosley (2009) \cite{Kasen2009} show how large amount of $^{56}$Ni can even extend the plateau. When the optically thick phase is over (i.e. the recombination front has receded all the way to the base of the envelope), $^{56}$Ni decay becomes the dominant power source and the light curve drops down to the $^{56}$Ni decay-powered tail.

\begin{figure}
\sidecaption
\includegraphics[width=\textwidth]{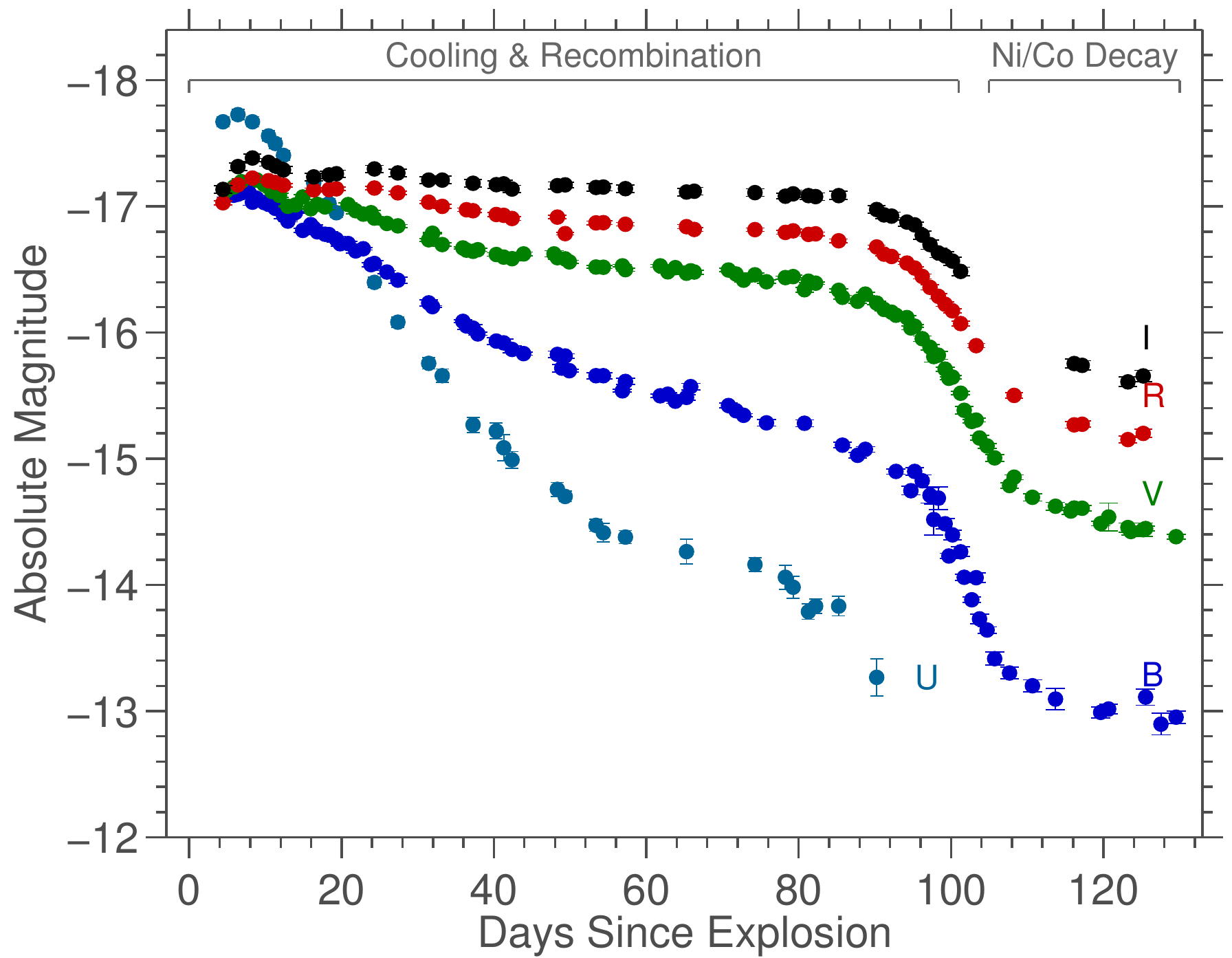}
\caption{Multi-band light curves of the Type IIP SN\,2013ab from \cite{Bose2015}. A $\sim100$-day plateau can be seen in the redder bands, while the blue bands decline due to increasing Fe-group line absorption. At roughly $100$ days, the photosphere recedes to the base of the (ejected) hydrogen envelope and the luminosity drops rapidly to the $^{56}$Ni-decay power tail. The bolometric light curve of the Type IIP SN\,1999em is presented in Figure \ref{fig:bols}. Plateau luminosities are observed in a large range of magnitudes ($-14$ to $-18$), but plateau lengths are typically tightly scattered around $100$ days.}
\label{fig:typical_iip}
\end{figure}

\subsubsection{The IIP/IIL Division}

Plateau luminosities are observed across a large range \cite{Patat1993, Patat1994}, from faint IIP's with a plateau magnitude of $-14$ to more luminous events at magnitude $-18$ (this is a factor of $\sim40$ in luminosity). The plateau lengths, on the other hand, are surprisingly uniform compared to the scatter in luminosities \cite{Arcavi2012a, Faran2014a, Anderson2014}. The scenario above intuitively suggests that the duration of the optically thick phase should depend on the mass of the hydrogen envelope of the star just before it exploded. Indeed, Popov (1993) \cite{Popov1993} finds that the plateau length, $t_p$ should depend mostly on the progenitor mass, $M$, with $t_p{\propto}\sqrt{M}$.

However, only a relatively narrow scatter of plateau lengths is observed, or no plateau at all (for SNe IIL), but nothing in between (i.e. no short plateaus of a few tens of days have been observed). Regarding a continuous single-population in light curve decline rates (from zero for IIP's to non-zero for IIL's), this is still under investigation. Arcavi et al. (2012) \cite{Arcavi2012a} and Faran et al. (2014) \cite{Faran2014} find a clear division between IIP and IIL light curves. Anderson et al. (2014) \cite{Anderson2014} and Faran et al. (2014) \cite{Sanders2015}, on the other hand, see a continuum of light curves (Fig. \ref{fig:iip_vs_iil}), though it's not yet clear if this is more consistent with one or possible two (overlapping) populations. 
Comparing these samples is not trivial for several reasons. The samples themselves are drawn from various surveys with different inherent selection biases and data quality. Also, the bands observed by Arcavi et al. (2012) \cite{Arcavi2012a} and Anderson et al. (2014) \cite{Anderson2014} behave differently ($R$-band vs. $V$-band; Fig \ref{fig:typical_iip}), though Faran et al. (2014) \cite{Faran2014a} see the IIP-IIL division in all bands. Additionally, the sample sizes are different, and the definitions used in the various works for parameterizing the light curves are not uniform. 

\begin{figure}
\sidecaption
\includegraphics[width=\textwidth,trim={30 0 60 0},clip]{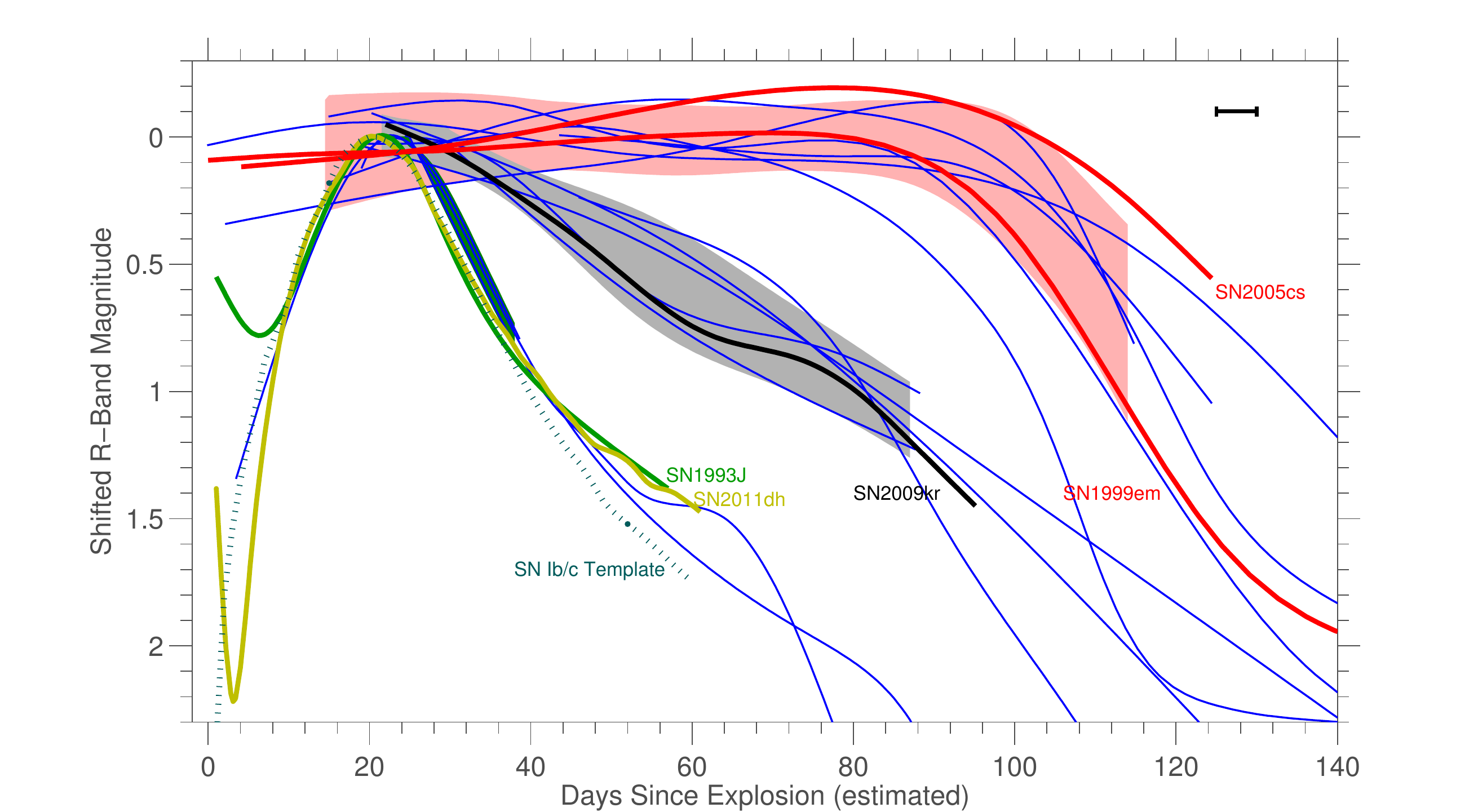}\\
\includegraphics[width=\textwidth]{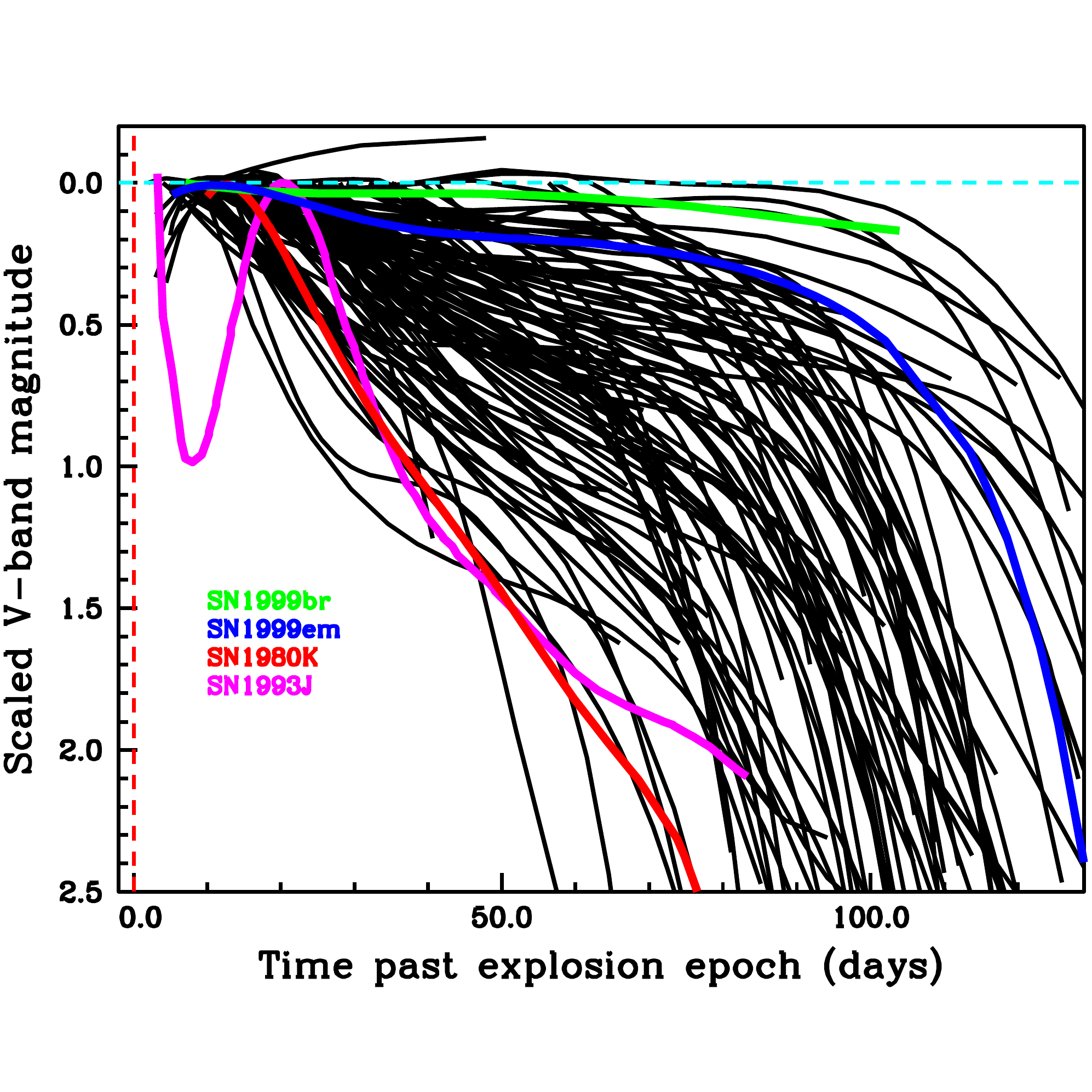}
\caption{Two separate populations, two overlapping populations or a single continuous one? Top: Noramlized Type II SN light curves in $R$-band from \cite{Arcavi2012a} (see also \cite{Faran2014a}) displaying distinct light curve populations for IIP, IIL and IIb light curves (with IIb being the most rapidly declining). Bottom: $V$-band from \cite{Anderson2014} (see also \cite{Sanders2015} displaying a continuum of light curve decline rates from IIP to IIL SNe.}
\label{fig:iip_vs_iil}
\end{figure}

An important clue regarding the nature of IIL SNe and their relation to IIP events is that IIL light curves also show a sudden drop (if observed long enough) as seen in the end of the optically thick phase of Type IIP light curves (Fig. \ref{fig:iil_breaks}) \cite{Valenti2015} (suggested earlier by Anderson et al. (2014) \cite{Anderson2014}, but their data were not as conclusive). This observation suggests that Type IIL SNe also go through an optically thick phase, except it is not at constant luminosity like in Type IIP's.  Intermediate objects between IIP and IIL events are thus not expected to look like IIP's with short plateaus, but rather IIP's with a decline (also in the redder bands) during their optically thick phase (i.e. that plateau perhaps gradually converts into a decline as one moves from IIP to IIL). 

All IIL SNe that have been observed long enough show such a drop, except for possibly SN\,1979C (until recently considered the prototypical IIL), though there is a gap in the data where its light curve drop might be. For IIL's where a drop is observed, it generally occurs slightly before the typical 100-day occurrence of Type IIP light curve drops.

\begin{figure}
\sidecaption
\includegraphics[width=\textwidth]{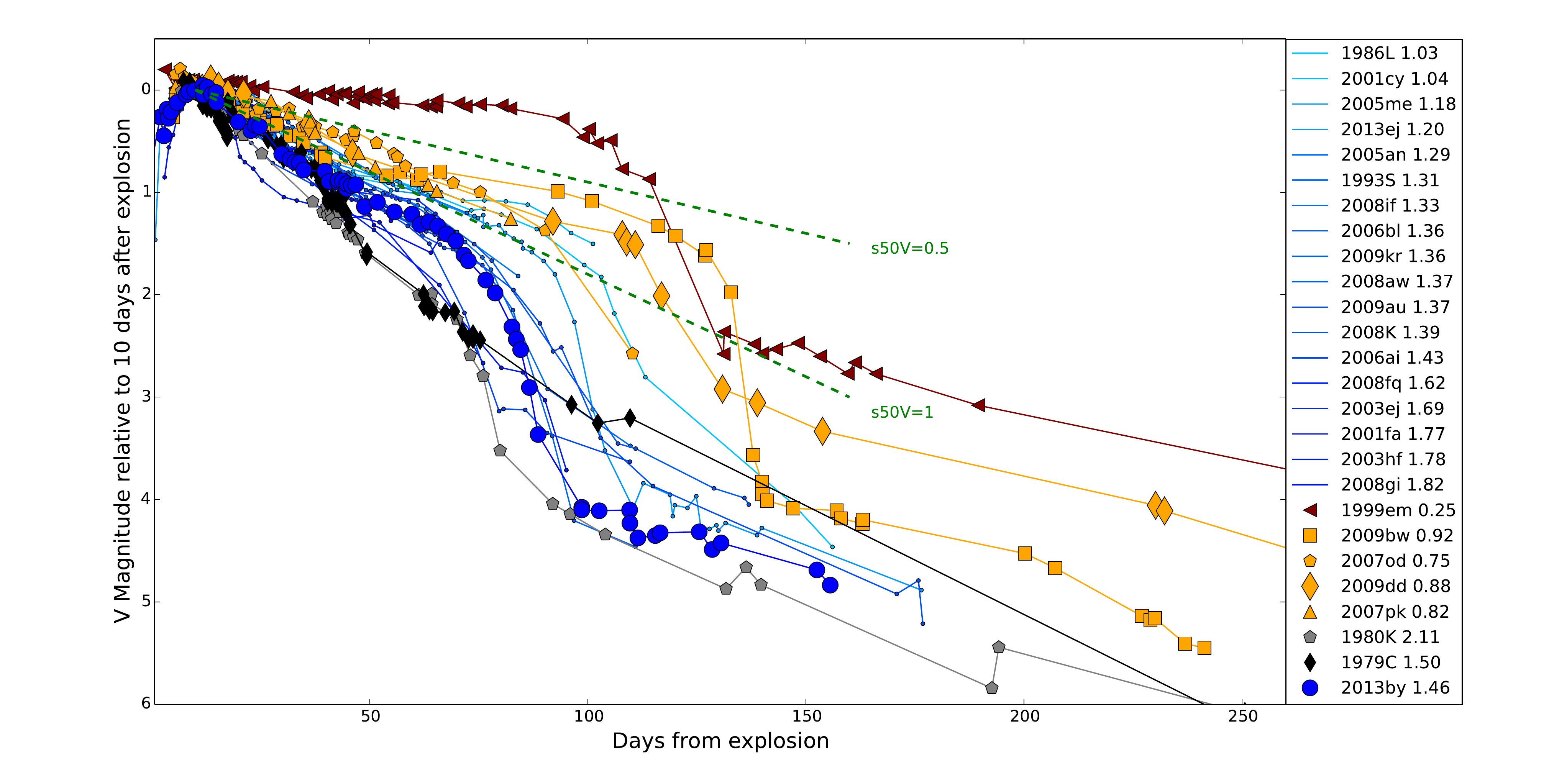}
\caption{V-band light curves of Type IIP/L SNe from \cite{Valenti2015} and references therein. The parameter s50V is the $V$-band decline rate (in magnitudes per $100$ days) measured between maximum light and $50$ days post explosion. The rapidly declining events (considered IIL's) shown here all have a ``drop'' indicative of a switch in power sources from cooling to radioactive decay, marking the existence of an optically thick phase like for IIP SNe. Curiously, SN\,1979C (until recently considered the prototypical IIL) may be the only event without such a drop. Figure from \cite{Valenti2015}.}
\label{fig:iil_breaks}
\end{figure}

Patat et al. (1994) \cite{Patat1994} and Li et al. (2011) \cite{Li2011} found that generally IIL SNe are brighter than IIP SNe. This was more recently confirmed in a correlation found by Anderson et al. (2014) \cite{Anderson2014} that SNe with higher light curve rates of decline during the optically thick phase are brighter at peak (Fig. \ref{fig:mmax_vs_slope}). Together with the Valenti et al. (2015) \cite{Valenti2015} result, this suggests that Type IIL's could be Type IIP's with an added luminosity component. This component declines with time and thus appears mostly during the optically thick phase. The nature of this added component is not yet clear. It may be from a larger initial progenitor radius for IIL's compared to IIP's, from interaction with a more massive CSM for IIL's, from power injected from a newly formed magnetar, or something else. One caveat to this picture is the shorter optically thick phase for IIL's compared to IIP's (Fig. \ref{fig:iil_breaks}), indicating there may still be something intrinsically different between these subclasses, in addition to the added luminosity component.

\begin{figure}
\sidecaption
\includegraphics[width=0.6\textwidth,trim={0 30 0 0},clip]{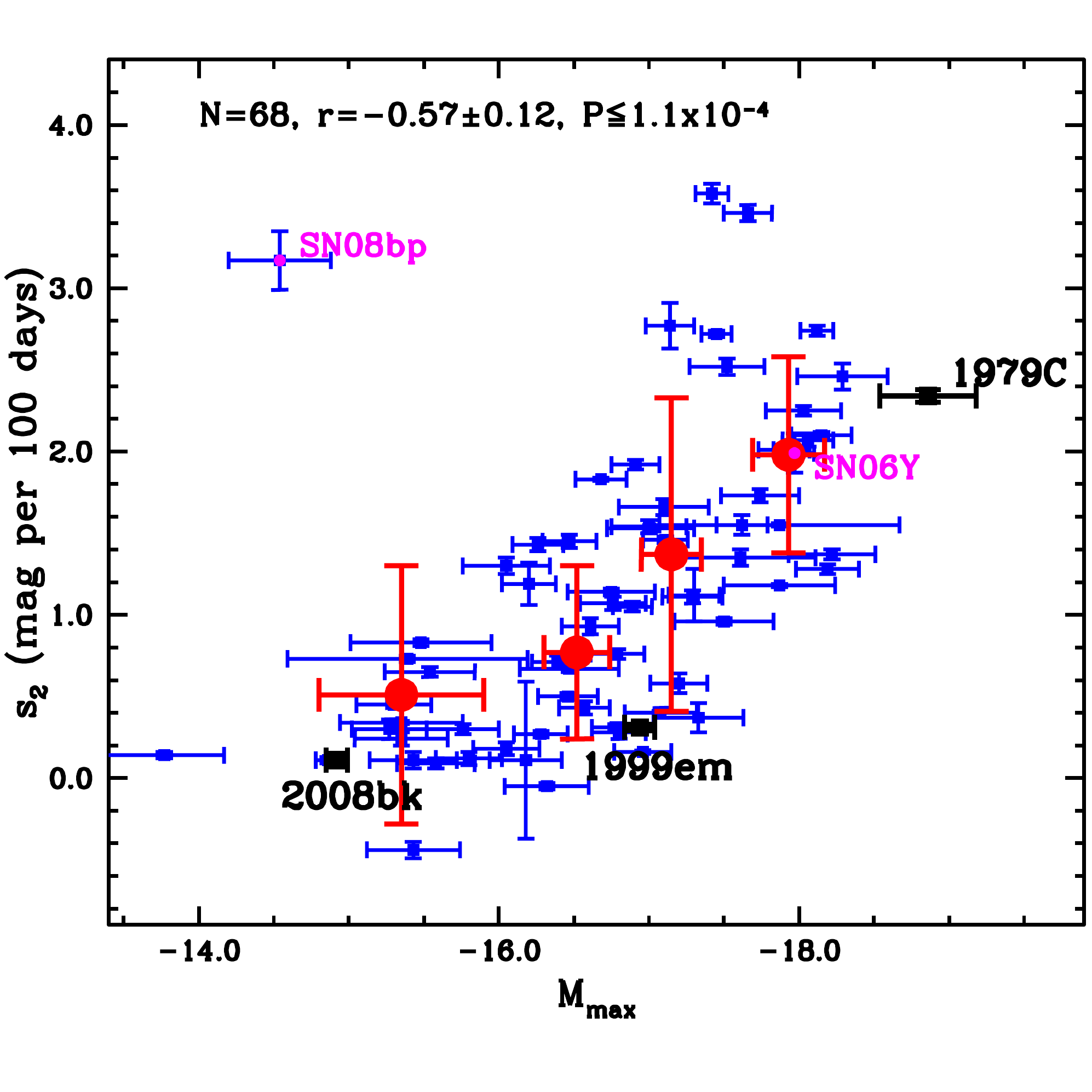}
\caption{$V$-band correlation between peak magnitude (denoted $\textrm{M}_{\textrm{max}}$) and decline rate (denoted $\textrm{s}_{\textrm{2}}$) for IIP/IIL events from \cite{Anderson2014}. Together with the finding that IIL SNe also show a late-time light curve ``drop'' indicating the switch form an optically thick to an optically thin phase, this correlation suggests that IIL SNe could simply be IIP events with an added declining luminosity component during the optically thick phase. The source of this added luminosity, however, is not clear. Figure from \cite{Anderson2014}.}
\label{fig:mmax_vs_slope}
\end{figure}

The origin of IIL SNe and their connection to IIP SNe remains a topic of debate. Homogeneously collected samples with consistent data sampling, multi-band coverage (allowing for bolometric light curves to be analyzed) and uniform parameterizations are key. 

Early-time observations are crucial for checking if a division into classes can be found in the rise-times of these events. Such studies have recently been undertaken \cite{Gall2015, Gonzalez-Gaitan2015, Rubin2016} but their results are still not conclusive.

\subsection{1987A-Like SNe}

As mentioned earlier, the progentior of SN\,1987A was a BSG, a type of star with a smaller radius than a RSG. Thus, when a BSG explodes, a large part of the energy carried by the shock goes into expanding the envelope of the star and less goes to heating and ionizing it. Therefore a plateau is not seen in the light curve of SN\,1987A, but instead it is dominated by $^{56}$Ni decay power. A brief shock cooling phase can be seen in the initial decline of the bluer bands (up to approximately day 20; Fig. \ref{fig:1987a}).

\begin{figure}
\sidecaption
\includegraphics[width=\textwidth]{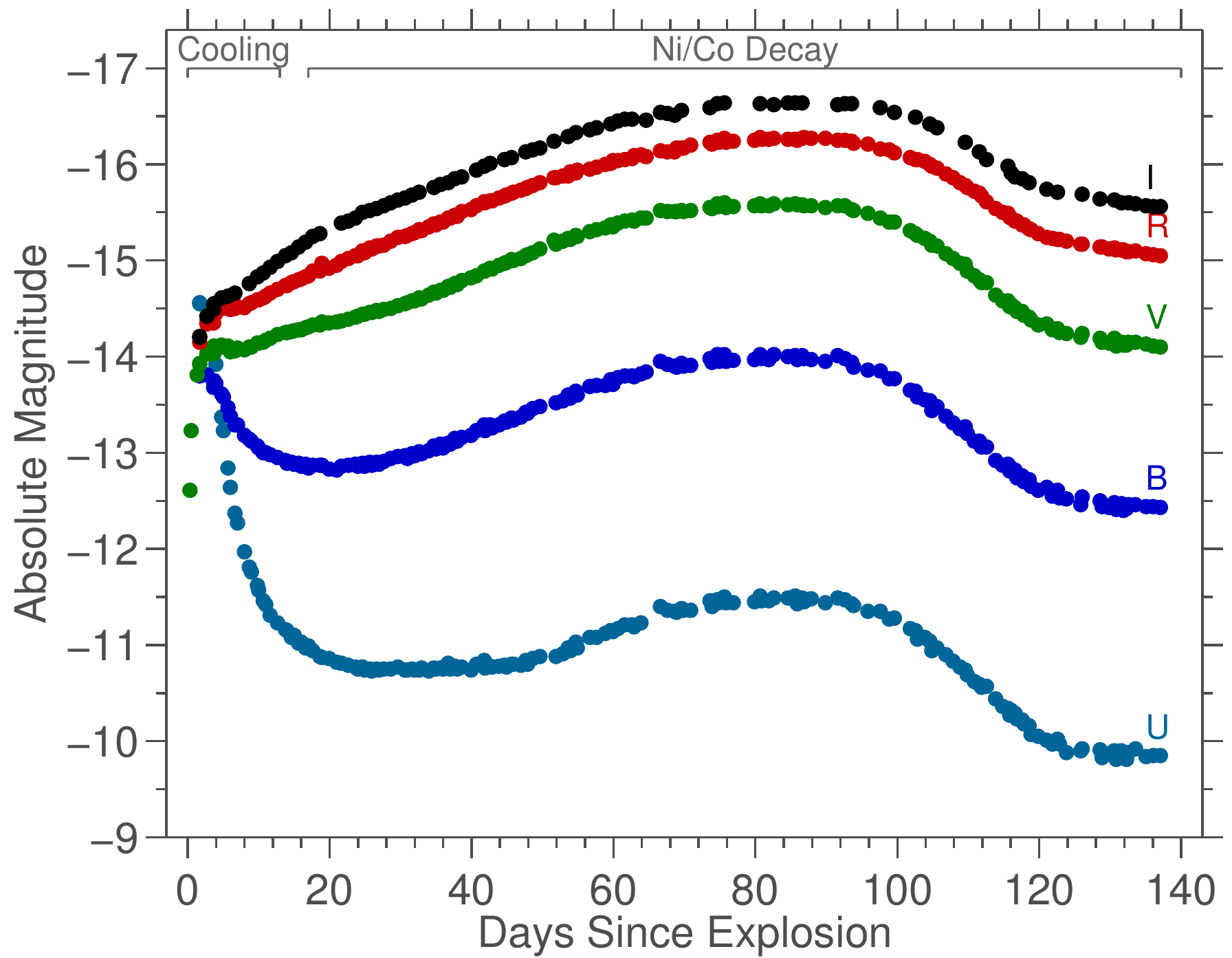}
\caption{Multi-band light curves of SN\,1987A from \cite{Hamuy1990}. All bands display a long ($\sim90$-day) rise to a peak. The $U$- and $B$-bands display an earlier peak and decline before the rise to the main peak, attributed to cooling of the expanding ejecta. The bolometric light curve of SN\,1987A is presented in Figure \ref{fig:bols}.}
\label{fig:1987a}
\end{figure}

Several other SN\,1987A-like events have since been identified \cite{Pastorello2005, Kleiser2011, Pastorello2012, Taddia2012, Taddia2016}, displaying some variability in their peak magnitudes and rise times (within the long rise-time definition of this class).

\begin{figure}
\sidecaption
\includegraphics[width=0.6\textwidth]{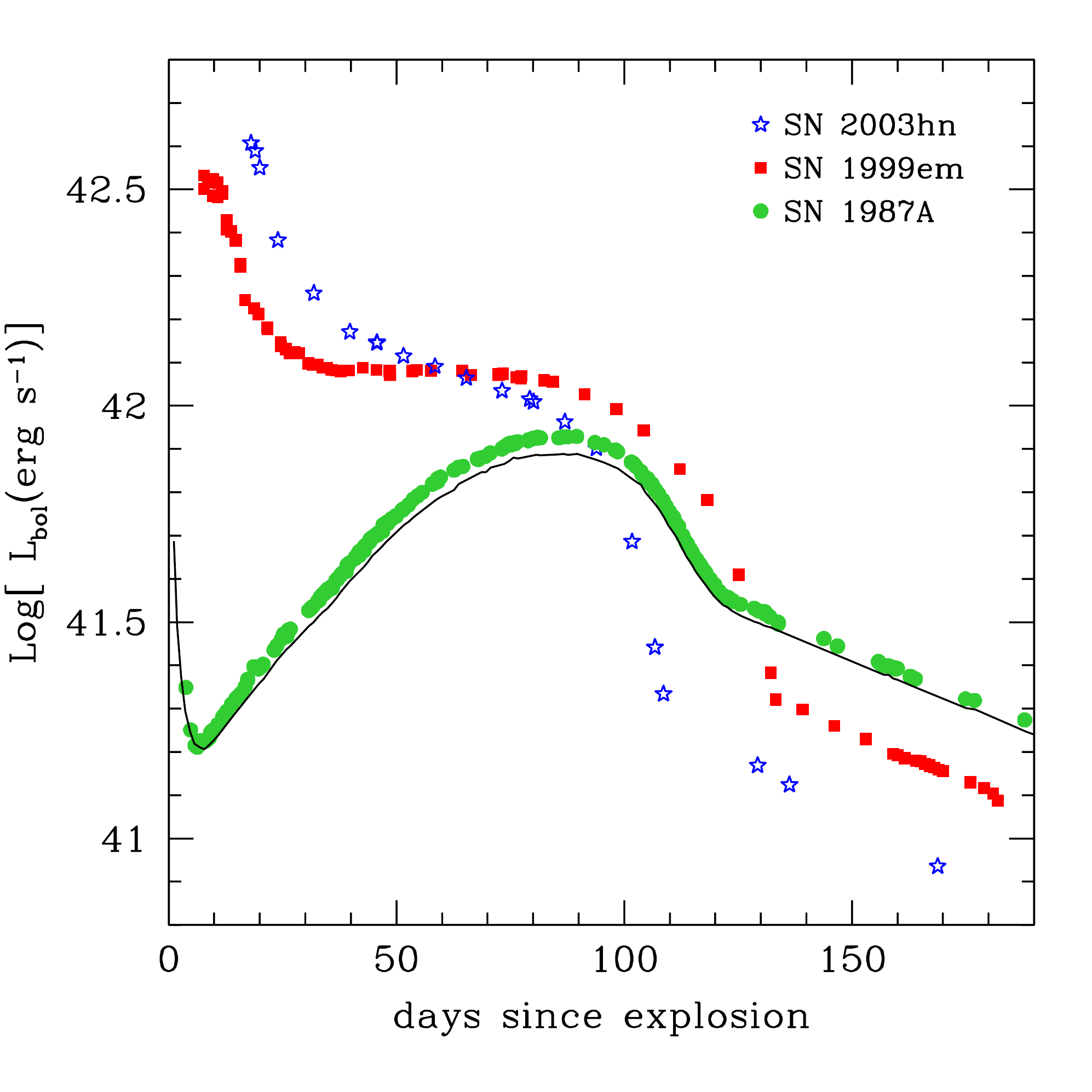}
\caption{Bolometric light curves from \cite{Bersten2009} (symbols) and \cite{Suntzeff1990} (solid line for SN\,1987A). Figure from \cite{Bersten2009}.}
\label{fig:bols}
\end{figure}

\subsection{IIb SNe}
\label{subsec:light_curves_iib}

Type IIb progenitors, having lost some or most of their hydrogen envelope before explosion, are also not capable of sustaining a light curve plateau. Instead they decline faster than IIP and IIL events following the Ni-powered peak (Fig. \ref{fig:iip_vs_iil}; top). In some events, a preceding light curve peak can be seen before the Ni-powered one, considered to be a shock cooling component (it is much shorter than the optically thick phase in IIP's). SN 1993J displayed a week-long shock cooling component (Fig. \ref{fig:1993j}), while SN\,2011dh displayed a similar light curve feature that lasted only $3$ days \cite{Arcavi2011}. 

\begin{figure}
\sidecaption
\includegraphics[width=\textwidth]{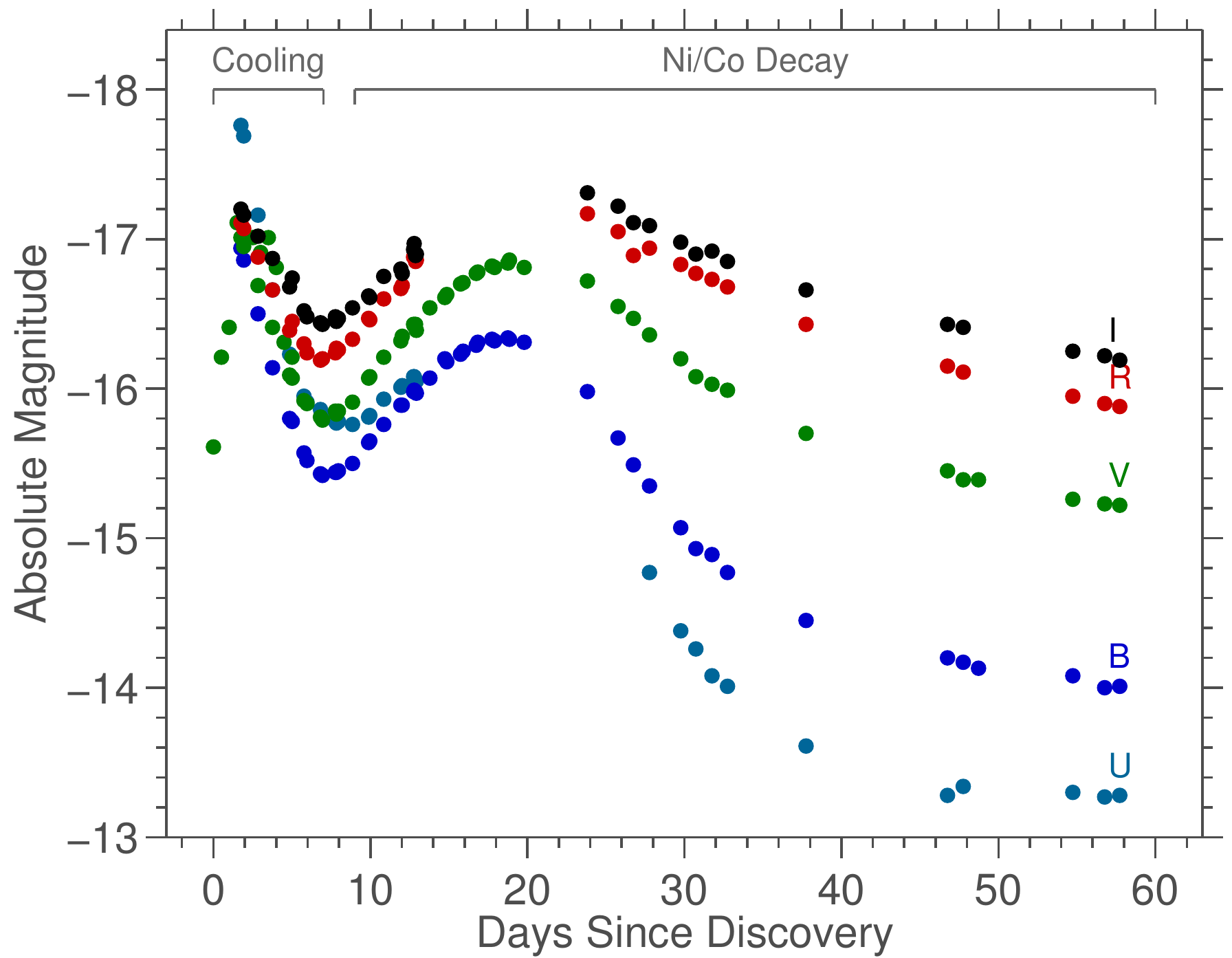}
\caption{Multi-band light curves of the Type IIb SN\,1993J from \cite{Richmond1994}. A double peak is seen in all filters (also the redder bands, unlike in SN\,1987A), implying the existence of a pre-explosion extended low-mass envelope, perhaps created through interaction of the progenitor with a binary companion.}
\label{fig:1993j}
\end{figure}

Unlike SN\,1987A, this shock cooling decline for IIb's is apparent also in the redder bands. This is an important difference, since the explosion of a star with a massive envelope is not expected to have a decline in the redder bands of its light curve during the envelope cooling phase \cite{Nakar2014}. To explain the early $R$-band decline seen in some IIb's, a non-standard density structure for the progenitor envelope needs to be evoked \cite{Woosley1994, Bersten2012, Nakar2014, Piro2015}. Specifically the progenitor star would need a low mass ($\lesssim1M_{\odot}$, even as low as $\sim0.01M_{\odot}$) but extended ($\sim10^{13}$\,cm) hydrogen envelope. The radius and mass of the core and those of the envelope are imprinted into the shock cooling part of the light curve (Fig. \ref{fig:iib_density_structure}; \cite{Bersten2012, Nakar2014, Piro2015}). Benvenuto, Bersten and Nomoto (2013) \cite{Benvenuto2013} show that such a density structure can be created through interaction with a binary companion.

\begin{figure}
\sidecaption
\includegraphics[width=0.6\textwidth]{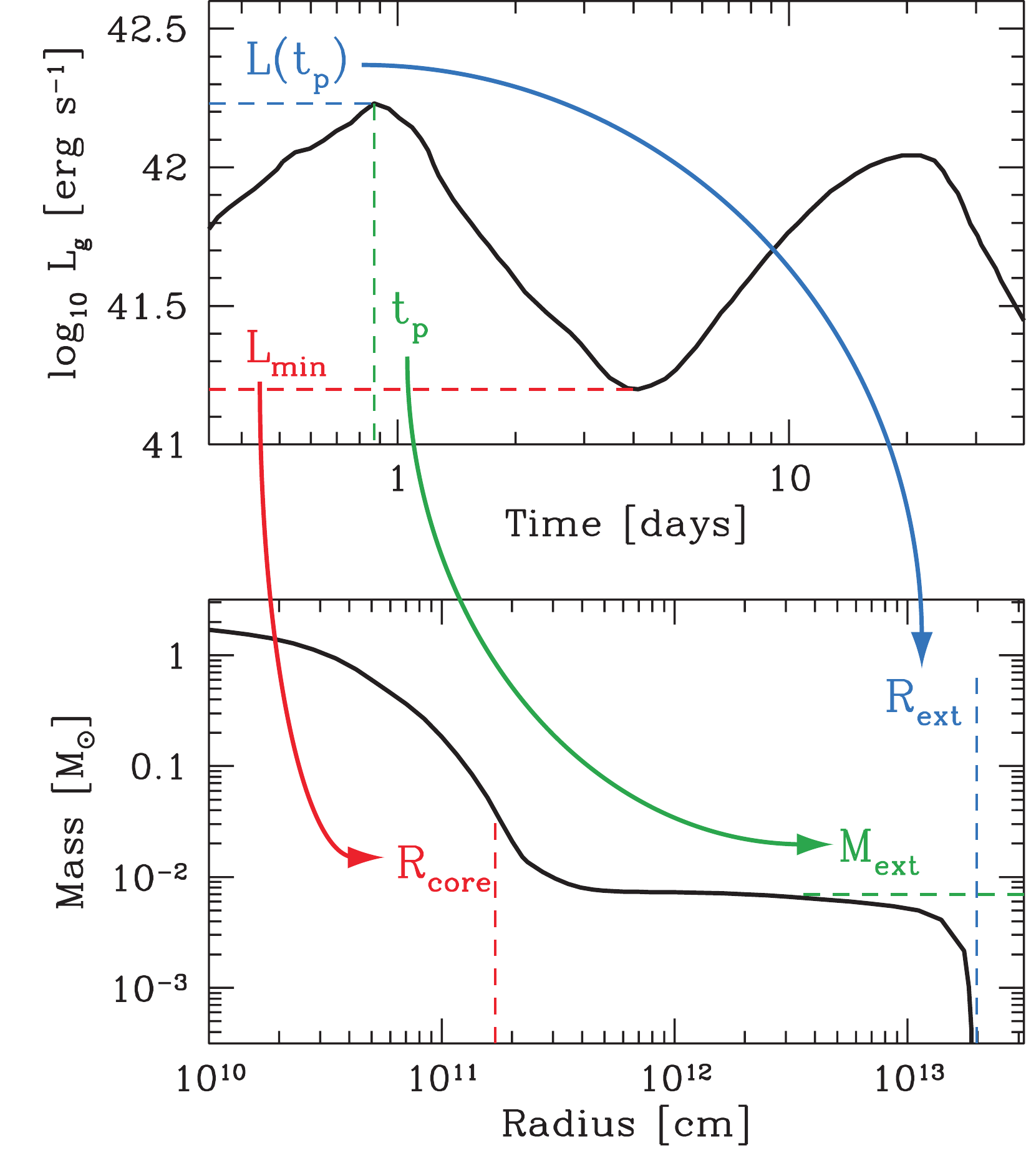}
\caption{A double peaked light curve (top), as seen for some Type IIb SNe, contains information about the peculiar density structure (involving a low mass extended envelope) of the progenitor (bottom). Such a progenitor density structure could be an indication of interaction with a binary companion prior to explosion. The time since explosion of the first light curve peak, $t_p$, its luminosity, $L\left(t_]\right)$, and the minimum luminosity between the two peaks, $L_{min}$, can be analytically connected to the mass in the extended envelope, $M_{ext}$, the radius of the extended envelope, $R_{ext}$, and the radius of the core, $R_{core}$, respectively. Figure from \cite{Nakar2014}.}
\label{fig:iib_density_structure}
\end{figure}

\subsection{IIn SNe}

Type IIn SNe produce the most luminous light curves of all H-rich events. This is explained by the added luminosity generated from CSM interaction. Type IIn's also display the widest range of luminosities (spanning approximately $6$ magnitudes) and light curve shapes (Fig. \ref{fig:iin_lcs}) of all H-rich SNe. This diversity is attributed to diversity and possibly asymmetry in the density structure and mass of the pre-explosion CSM. The luminosity and shape of the light curve thus contain information about the CSM which can be traced to the mass-loss history of the progenitor before exploding. This is true also for the radio light curves \cite{Kotak2006a}.

Many Type IIn progenitors brighten significantly before explosion, in so-called SN precursor events (e.g. \cite{Ofek2013}). Such precursors are common for IIn SNe: over $50\%$ of them have at least one pre-explosion outburst brighter than $3\cdot10^7L_{\odot}$ within $4$ months prior to explosion \cite{Ofek2014}. These precursors are interpreted as violent ejections of several solar masses of material possibly due to pre-explosion instabilities in the interior of the star \cite{Quataert2012}. Type IIn light curves and their implications for the mass loss from massive stars is discussed in more detail in ``Interacting Supernovae''. 

\begin{figure}
\sidecaption
\includegraphics[width=0.5\textwidth]{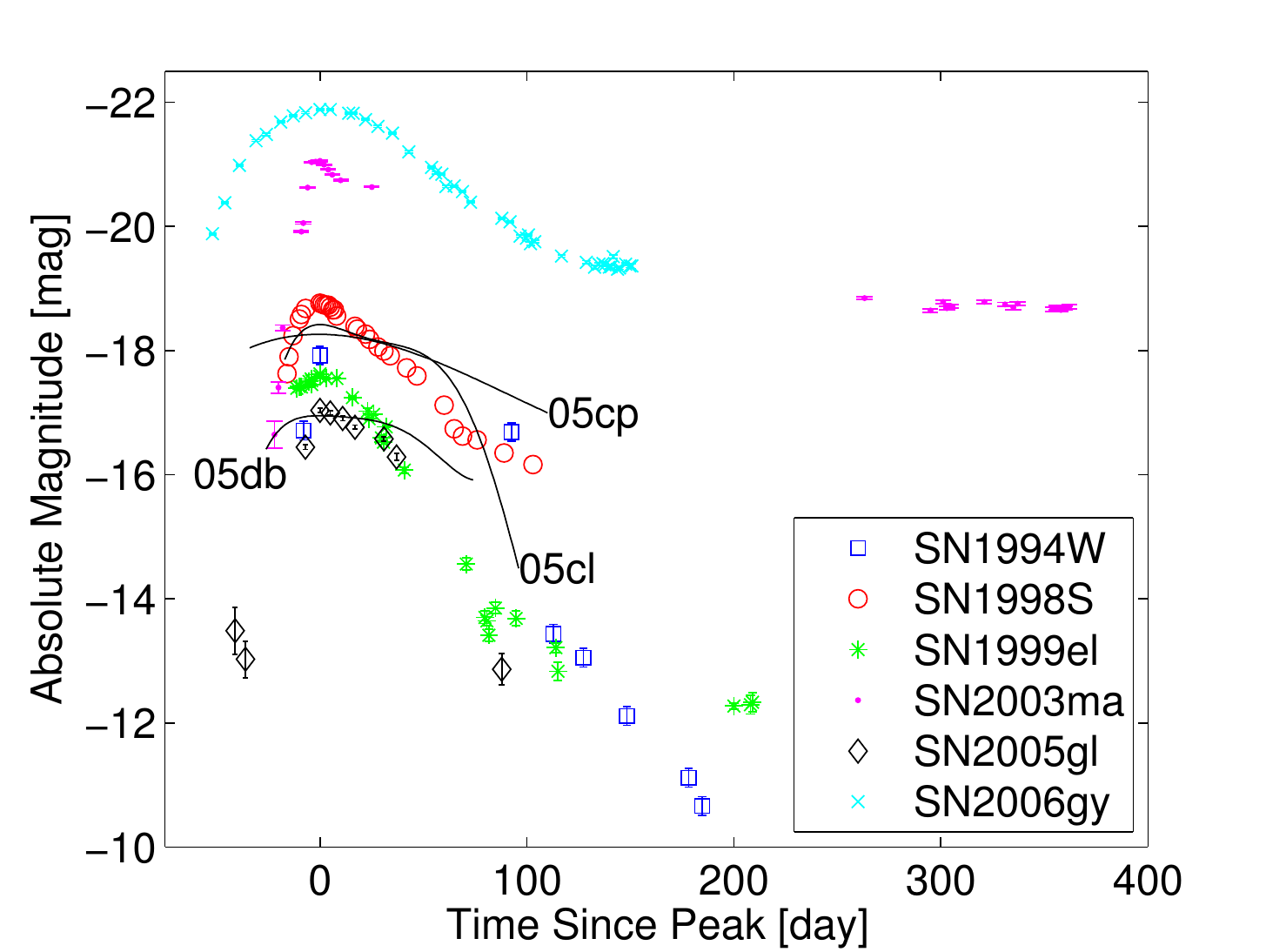}\includegraphics[width=0.5\textwidth]{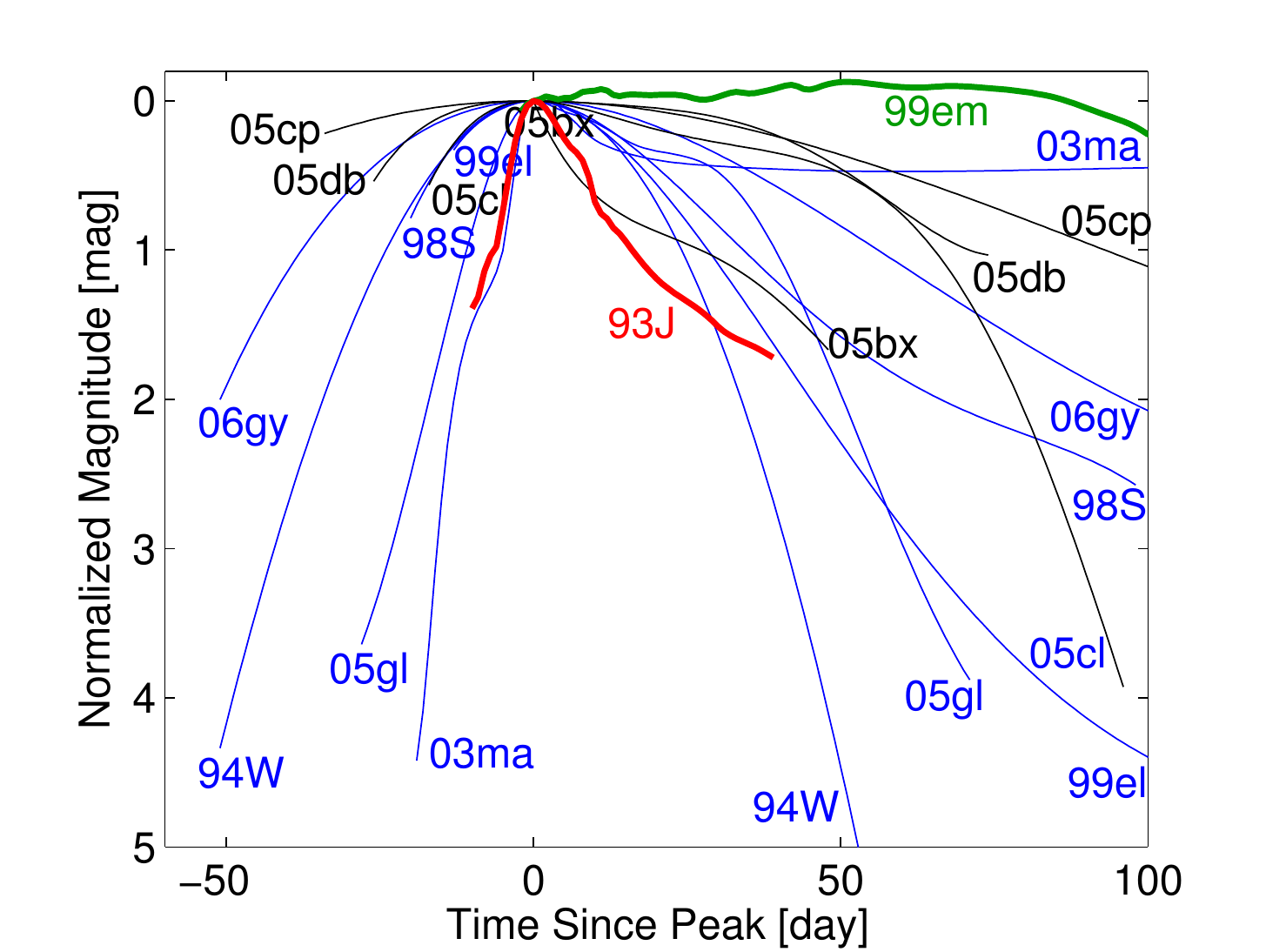}
\caption{Type II SN light curves from \cite{Kiewe2012} and references therein, in absolute magnitudes (left) and normalized to peak magnitude (right). Type IIn SN light curve peaks span almost $6$ magnitudes (a factor of $\sim250$ in luminosity), and display a large range of rise and decline rates between the rapidly declining Type IIb SN\,1993J and the plateau SN\,1999em (both shown for comparison in the right plot). The large diversity in light curves is likely related to the large diversity in CSM mass and density structure surrounding the progenitors of Type IIn SNe. Figures from \cite{Kiewe2012}.}
\label{fig:iin_lcs}
\end{figure}

\section{Spectra}
\label{sec:spectra}

As mentioned above, SN ejecta are initially very hot, then cool as they expand. They also become transparent, and the photosphere slowly recedes (in mass) through the expanding ejecta. This causes the spectra to evolve with time from the early (few days from explosion) stage of a featureless blackbody to the intermediate (few weeks) so-called ``photospheric'' stage to the late (few months) ``nebular'' stage. Sometimes, H-rich SN spectra will display narrow high-ionization features in the first few hours from explosion.

\subsection{Hours From Explosion: The Flash Ionization Phase}

If caught at very early times, H-rich SN spectra can show narrow features of high ionization species \cite{Niemela1985,Gal-Yam2014,Shivvers2015,Khazov2016}. These features promptly disappear within hours to days. The interpretation is that these features are the response of extended progenitor winds or other CSM to the ionizing shock-breakout flash. Depending on the location of this CSM, it can be swept up by the SN ejecta within hours to days from explosion. As such, these ``flash spectra'' features can briefly reveal the recent mass loss history of the progenitor in the narrow time window between the collapse of the core and the ejecta sweeping up the immediate stellar surroundings.

\begin{figure}
\sidecaption
\includegraphics[width=\textwidth]{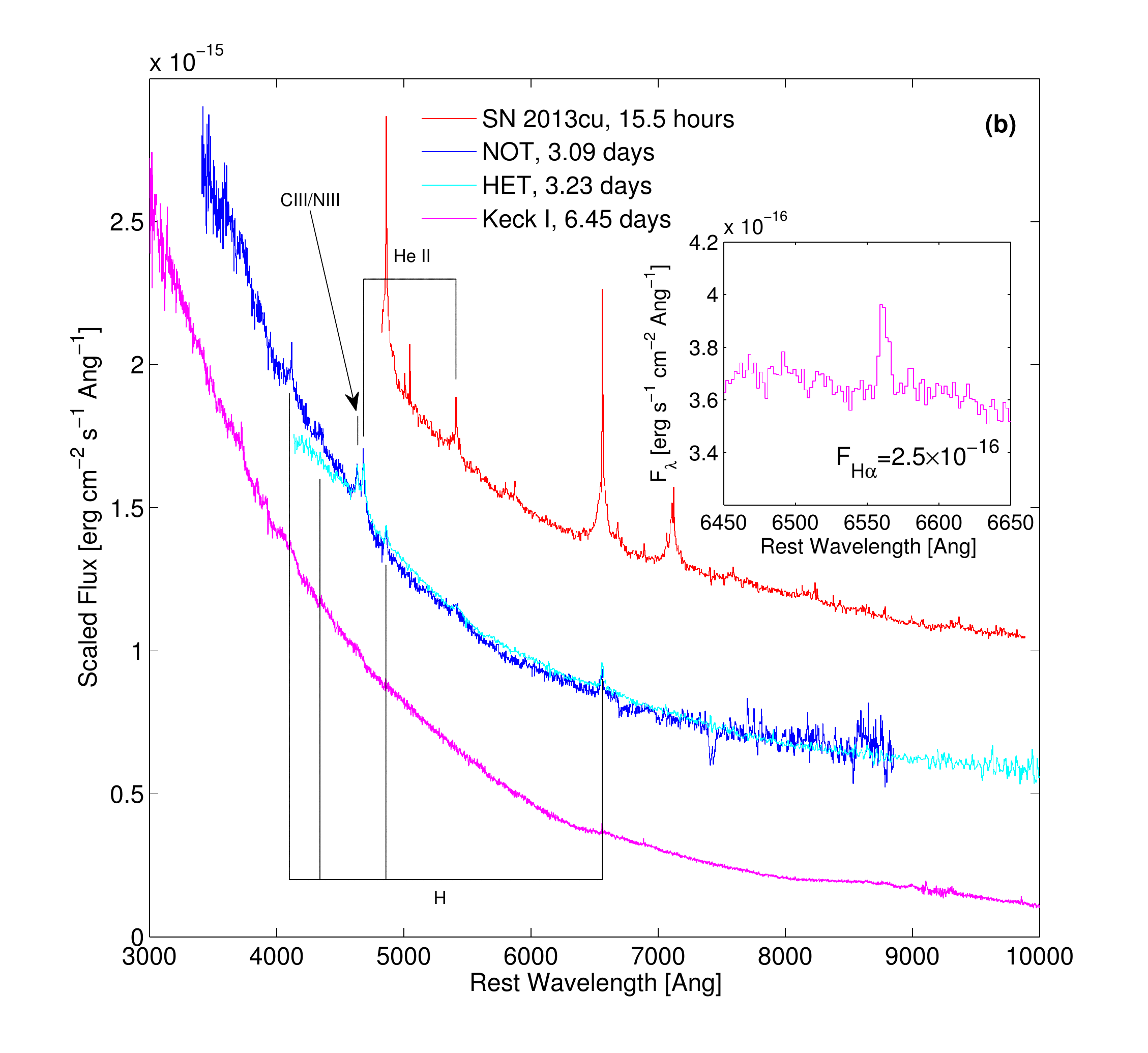}
\caption{Early spectra of the Type IIb SN\,2013cu from \cite{Gal-Yam2014}. The first spectrum (top) is taken only $\approx15.5$ hours after explosion and shows ``flash spectroscopy'' signatures - narrow emission lines from CSM immediately adjacent to the progenitor. These lines weaken substantially within a few days as the ejecta clear out the region from which the lines originated. A few days post-explosion the spectrum is a nearly featureless blue continuum, as seen for most all H-rich SNe at this stage.}
\label{fig:flash}
\end{figure}

\subsection{Days From Explosion: The Shock Cooling Phase}

For large progenitors, much of the SN shock energy goes to heating the ejecta, which then take a few days to cool, whereas for compact progenitors much of the energy goes to expansion and they cool more quickly. The rate at which the ejecta cool during the first days following explosion can thus be used to infer the radius of the progenitor (e.g. \cite{Nakar2010, Rabinak2011}). It is therefore very beneficial to obtain spectra or at least color information during the first days of a SN (e.g. Fig. \ref{fig:cooling}).  

\begin{figure}
\sidecaption
\includegraphics[width=\textwidth,trim={0 197 0 0},clip]{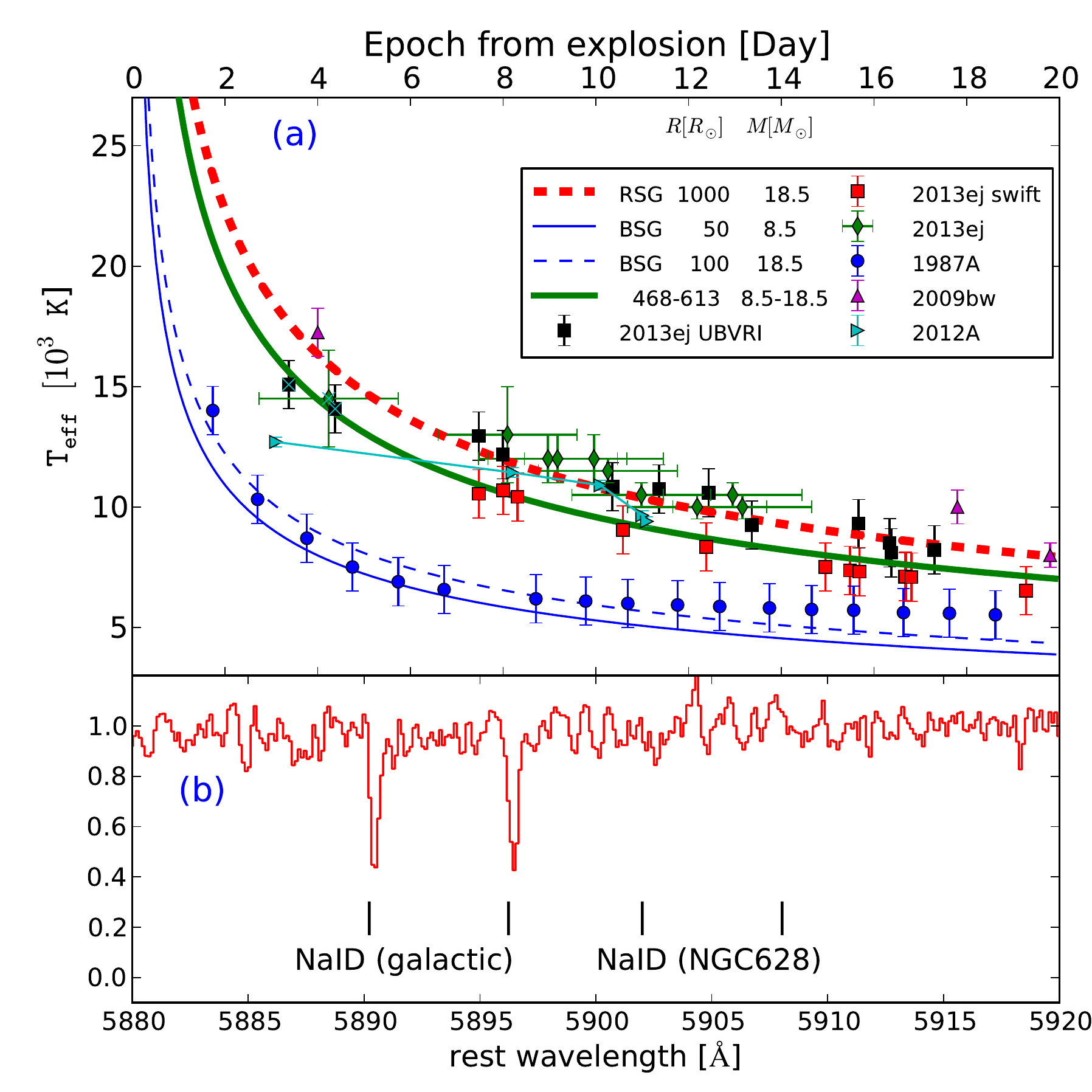}
\caption{Cooling of Type II SN ejecta as measured from blackbody continuum effective temperature fits (from \cite{Valenti2013} and references therein). SN\,1987A cooled faster than normal IIP SNe because its progenitor was more compact. Models from \cite{Rabinak2011} models (separated to RSGs and BSGs) can be used to constrain the radius of H-rich SN progenitors from such temperature measurements.}
\label{fig:cooling}
\end{figure}

Spectra of Type II SNe obtained during the start of the cooling phase tend to be featureless high-temperature (few $10^4$\,K) blackbody curves such as the bottom spectrum in Figure \ref{fig:flash}. 

\subsection{Weeks From Explosion: The Photospheric Phase}

As the ejecta cool, most Type II SN spectra develop broad (few $10^3-10^4$\,km\,s$^{-1}$) P-Cygni features \cite{Kirshner1973}. The Fe-group lines could possibly be used to deduce the metallicity of the progenitor star \cite{Dessart2014}.

Type IIL SNe have relatively lower absorption to emission ratios in their H$\alpha$ profile compared to Type IIP's (first noticed by Patat et al. (1994) \cite{Patat1994} and most recently quantified by Gutierrez et al. (2014) \cite{Gutierrez2014}; Fig. \ref{fig:specs_iip_iil}). These differences are attributed to possible different envelope masses, density profiles or CSM properties between IIP and IIL progenitors \cite{Schlegel1996}. A lower envelope mass would produce less absorption overall, a steeper density profile would produce less absorption at high velocities (thus not producing a well-defined P-Cygni profile), and more massive CSM could scatter additional light that would fill in the absorption feature. 

Type IIn SNe display different profiles altogether. The Balmer series is seen in narrow (few $10^2$\,km\,s$^{-1}$) emission with an intermediate ($\sim10^3$\,km\,s$^{-1}$) and/or broad (few $10^3$\,km\,s$^{-1}$) base (e.g. \cite{Kiewe2012}; Fig. \ref{fig:specs_phot}), giving the line a Lorenzian-like profile. The broad base seen here (and occasionally also in the flash ionization spectra) may be influenced by electron scattering wings and not actual kinetic motion of bulk material (e.g. \cite{Groh2014, Dessart2016}). 

\begin{figure}
\sidecaption
\includegraphics[width=\textwidth]{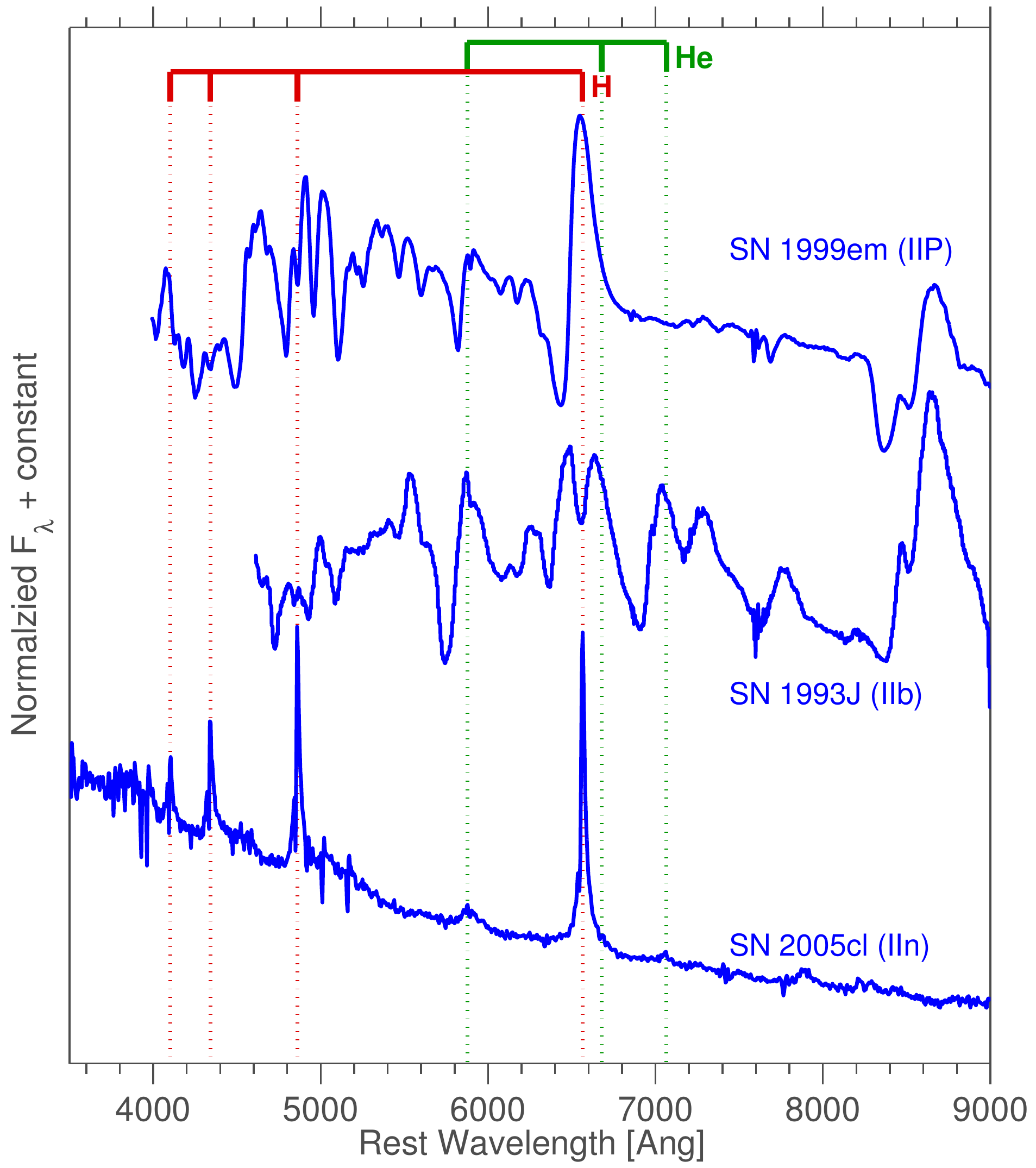}
\caption{Spectra of the Type IIP SN 1999em (from \cite{Leonard2002}), the Type IIb SN 1993J (from \cite{Barbon1995}) and the Type IIn SN 2005cl (from \cite{Kiewe2012}). Hydrogen lines (H$\alpha$ $6563{\textrm{\AA}}$, H$\beta$ $4861{\textrm{\AA}}$, H$\gamma$ $4341{\textrm{\AA}}$ and H$\delta$ $4102{\textrm{\AA}}$; marked in red) define all Type II SNe, with broad P-Cygni profiles in IIP's and IIb's but with narrow Lorenzian profiles in IIn's. Helium ($5876{\textrm{\AA}}$, $6678{\textrm{\AA}}$ and $7065{\textrm{\AA}}$; marked in green) broad P-Cygni features define the Type IIb class (the $5876{\textrm{\AA}}$ absorption feature overlaps with the Na I D $5889{\textrm{\AA}}$ and $5895{\textrm{\AA}}$ doublet; the He $6678{\textrm{\AA}}$ absorption features overlaps with the H$\alpha$ emission feature). SN\,1987A-likes display similar spectra as Type IIP's. Type IIL SNe display weaker absorption in the H$\alpha$ P-Cygni profile (Fig. \ref{fig:specs_iip_iil}).}
\label{fig:specs_phot}
\end{figure}

\begin{figure}
\sidecaption
\includegraphics[width=0.5\textwidth]{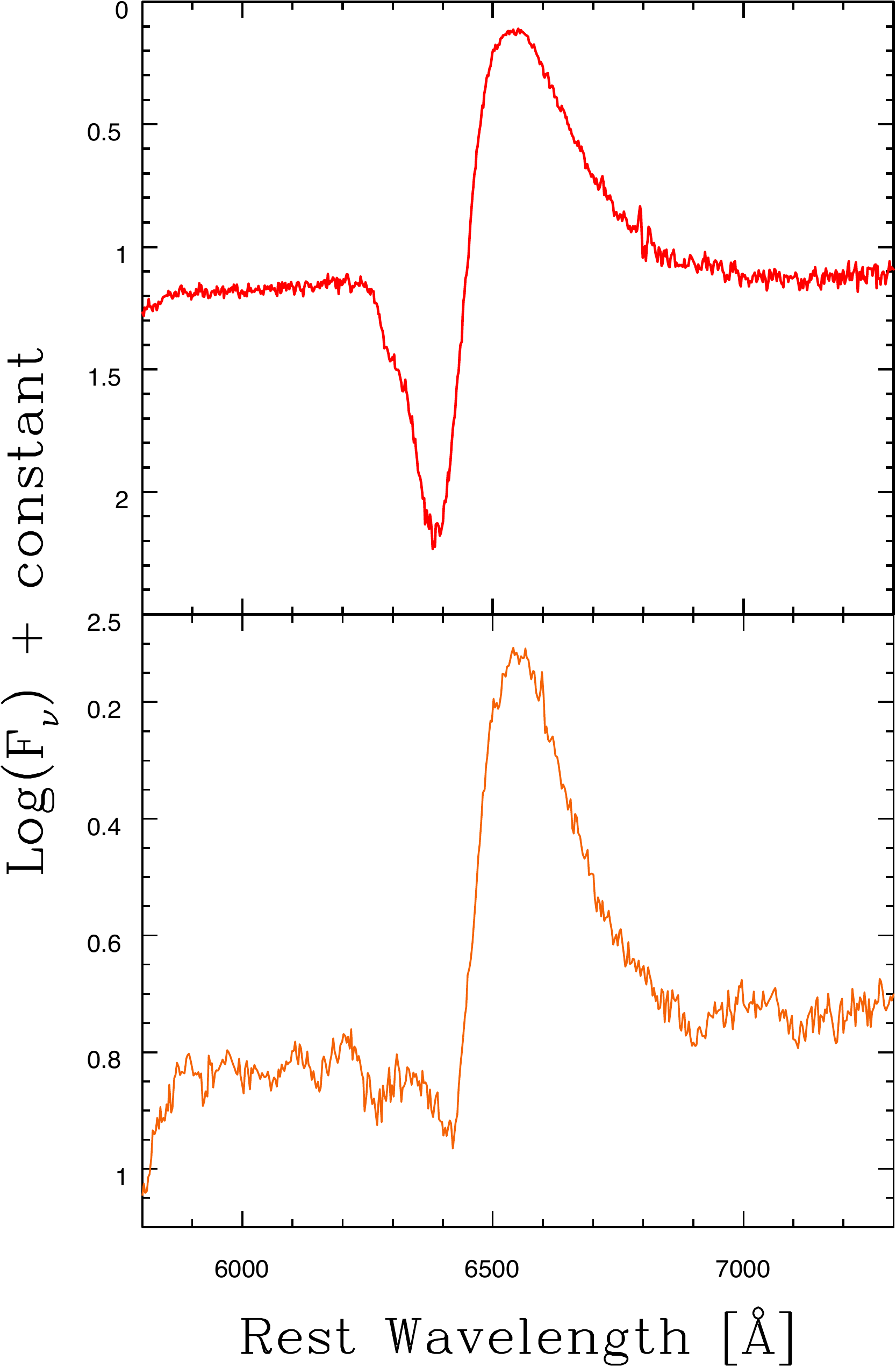}
\caption{H$\alpha$ profiles of the Type IIP SN\,2009bz (top) and the Type IIL SN\,2008aw (bottom) from \cite{Gutierrez2014}. This is an example of a broader correlation between the ratio of the absorption to emission of the H$\alpha$ feature and the slope of the light curve. Type IIP SNe show deeper P-Cygni absorption compared to Type IIL events. Figure adapted from \cite{Gutierrez2014}.}
\label{fig:specs_iip_iil}
\end{figure}

\subsubsection{Type IIP Velocity Evolution}

Each mass element in the SN ejecta is considered to be moving at a constant velocity $v_0$ that is proportional to its initial radius $r_0$, so that the outer ejecta is moving faster than the inner ejecta. Such expansion is known as homologous expansion. This allows velocity to be used as a coordinate that is co-moving with the ejecta. As the photosphere recedes inwards in the mass coordinate, it is also receding in the velocity coordinate. Observationally, this is seen as the gradual narrowing of the P-Cygni profiles with time. Note that this is not a deceleration of the ejecta, but the change in velocity sampled by the receding photosphere as it moves from the high velocity outer ejecta to the lower velocity inner ejecta.

Interestingly, the observed photospheric velocity evolution is quite uniform for Type IIP SNe (the measured velocities evolve approximately as $v{\propto}t^{-0.5}$; e.g. \cite{Nugent2006}). Potentially, this observed velocity evolution can be used to determine the phase of a Type II SN from a single spectrum. However, the normalization factor for the velocity evolution is different for different SNe (i.e. the velocity curves have the same shape but are shifted for different events; Fig. \ref{fig:velocities}).

\begin{figure}
\sidecaption
\includegraphics[width=0.6\textwidth]{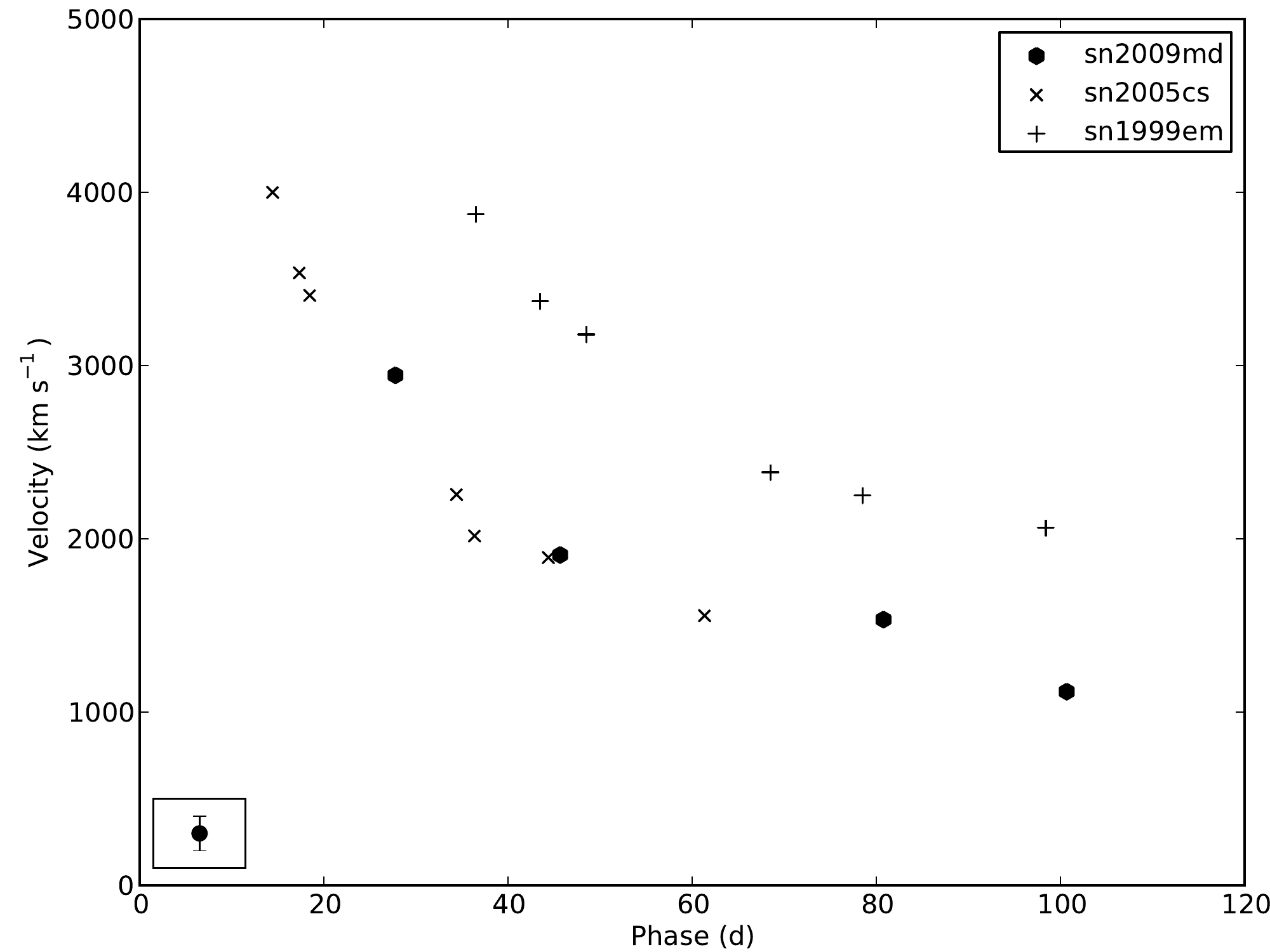}\,\,\includegraphics[width=0.39\textwidth,trim={370 50 0 0},clip]{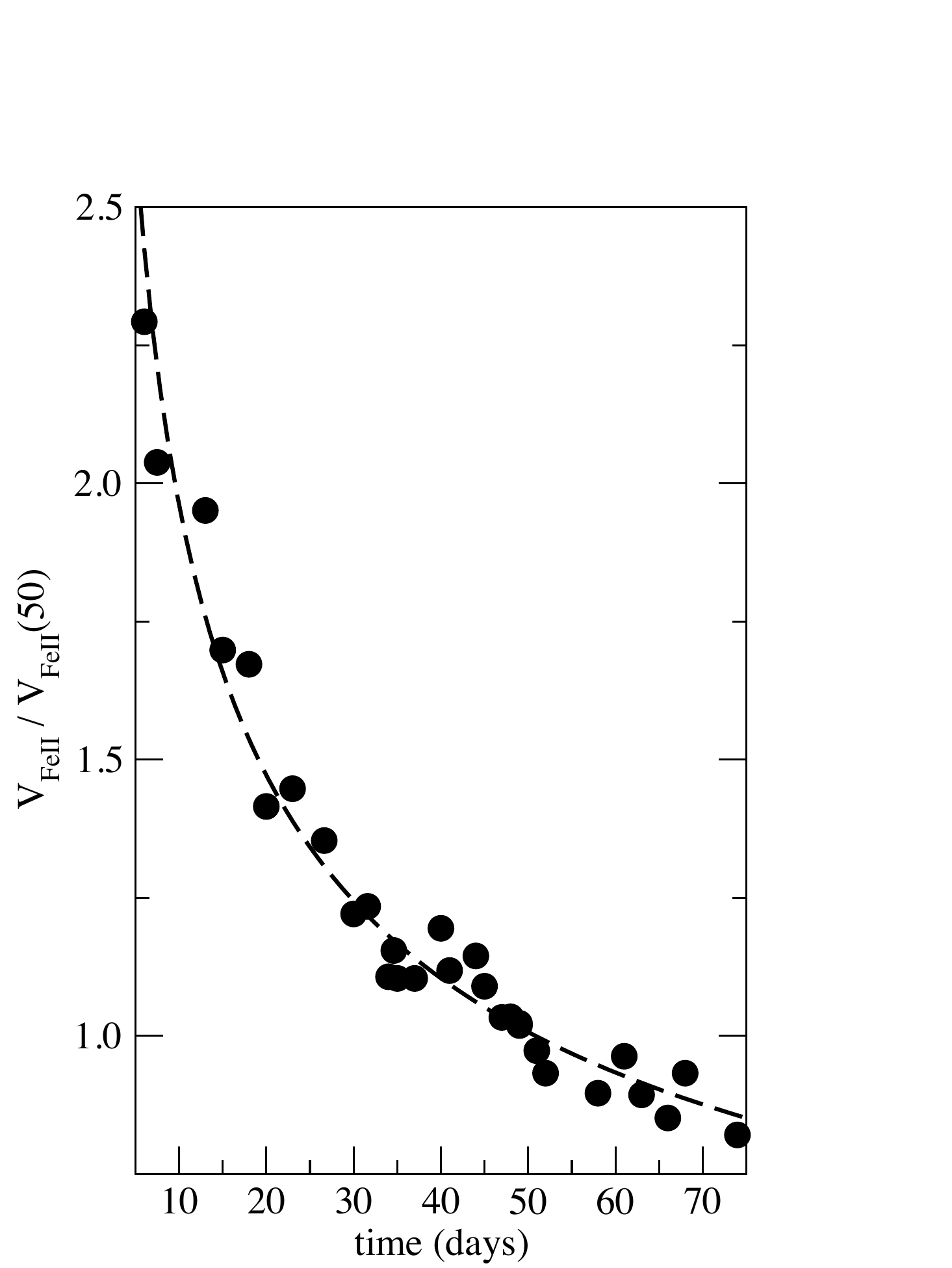}
\caption{Left: Velocities of the Sc II $6246{\textrm{\AA}}$ line for a few Type IIP SNe from \cite{Fraser2011} and references therein. The velocity can be seen to decrease with time as the photosphere recedes to slower-moving ejecta. SNe with more luminous plateaus (like SN\,1999em) display overall higher velocities than SNe with less luminous plateaus (like SNe 2005cs and 2009md). The bottom-left inset shows the typical measurement error for these data. Right: Normalized velocities of the Fe II $5169{\textrm{\AA}}$ line for various SNe from \cite{Nugent2006}. Once normalized, all velocities seem to lie on a single power-law evolution curve ($t^{-0.464}$ for this dataset). Left plot from \cite{Fraser2011}; Right plot from \cite{Nugent2006}.}
\label{fig:velocities}
\end{figure}

Expansion velocities are often measured (in deredshifted spectra) as the difference between the P-Cygni absorption minimum compared to the rest-wavelength of the measured line. Different elements display different velocities at a given time. This is because there is an effective photosphere for each element that is at a different velocity coordinate. Generally, lighter elements are concentrated in the outer layers of the progenitor and thus have the highest velocities, whereas heavier elements, being more internal, have lower velocities.

We mentioned that the normalization factor for the velocity evolution is different for different events. This factor, in fact, correlates with the absolute luminosity of the SN during the plateau \cite{Hamuy2002}. More luminous Type IIP SNe have larger normalization factors (higher velocities) than less luminous ones (Fig. \ref{fig:iip_cosmology}; most commonly the Fe II $5169{\textrm{\AA}}$ line is used for the velocities, and both the normalization factor and plateau luminosity are defined at day $50$ after explosion). This correlation lead Hamuy and Pinto (2001) \cite{Hamuy2002} to propose Type IIP SNe as cosmological standardizable candles. The method was refined by Nugent et al. (2006) \cite{Nugent2006} and later by Poznanski et al. (2009) \cite{Poznanski2009} by simultaneously fitting for extinction using the observed colors during the plateau. 

\begin{figure}
\sidecaption
\includegraphics[width=0.6\textwidth,trim={50 150 50 150},clip]{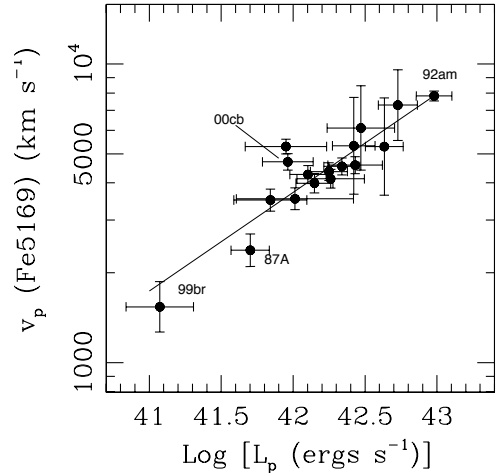}
\caption{Type IIP SN velocities of the Fe II $5169{\textrm{\AA}}$ line vs. bolometric luminosities, both measured at day $50$ (middle of the plateau) from \cite{Hamuy2002}. This correlation allows Type IIP SNe to be used as standard candles for distance measurements.}
\label{fig:iip_cosmology}
\end{figure}

Other methods for using Type IIP events as standard candles exist. The ``Expanding Photosphere Method'' (EPM; \cite{Kirshner1975, Eastman1996}) and the ``synthetic spectral atmosphere  fitting''  method  (e.g. \cite{Baron2004, Dessart2006}) have the advantage of being rooted in theoretical modelling of the explosions (rather than an empirical relation as the one described above). However, both of the theory-driven methods require high signal-to-noise spectra and photometry, in order to allow accurate comparison to the models, as well as early-time observations to precisely determine the explosion date. This makes the methods challenging for implementation among high-redshift SNe.

\subsubsection{Type IIP High Velocity H$\alpha$ or Si II Feature}

Some Type IIP SNe display an absorption feature just bluewards of the H$\alpha$ absorption (Fig. \ref{fig:hvha}). Two common identification of this feature exist: that it is a disjoint high velocity component of H$\alpha$, or that it is Si II absorption, more commonly seen in Type Ia SNe (see ``Type Ia Supernovae''). 

The high velocity H$\alpha$ option can be interpreted as a signature of interaction between the ejecta and the CSM \cite{Chugai2007}. Specifically, the ``normal'' velocity H$\alpha$ originates from the receding photosphere, while the high velocity component is generated further out, in otherwise-transparent material. If the outer material is interacting with the CSM, then shock waves might excite the hydrogen in the outer layers, causing a second, high-velocity absorption feature.

\begin{figure}
\sidecaption
\includegraphics[width=0.6\textwidth]{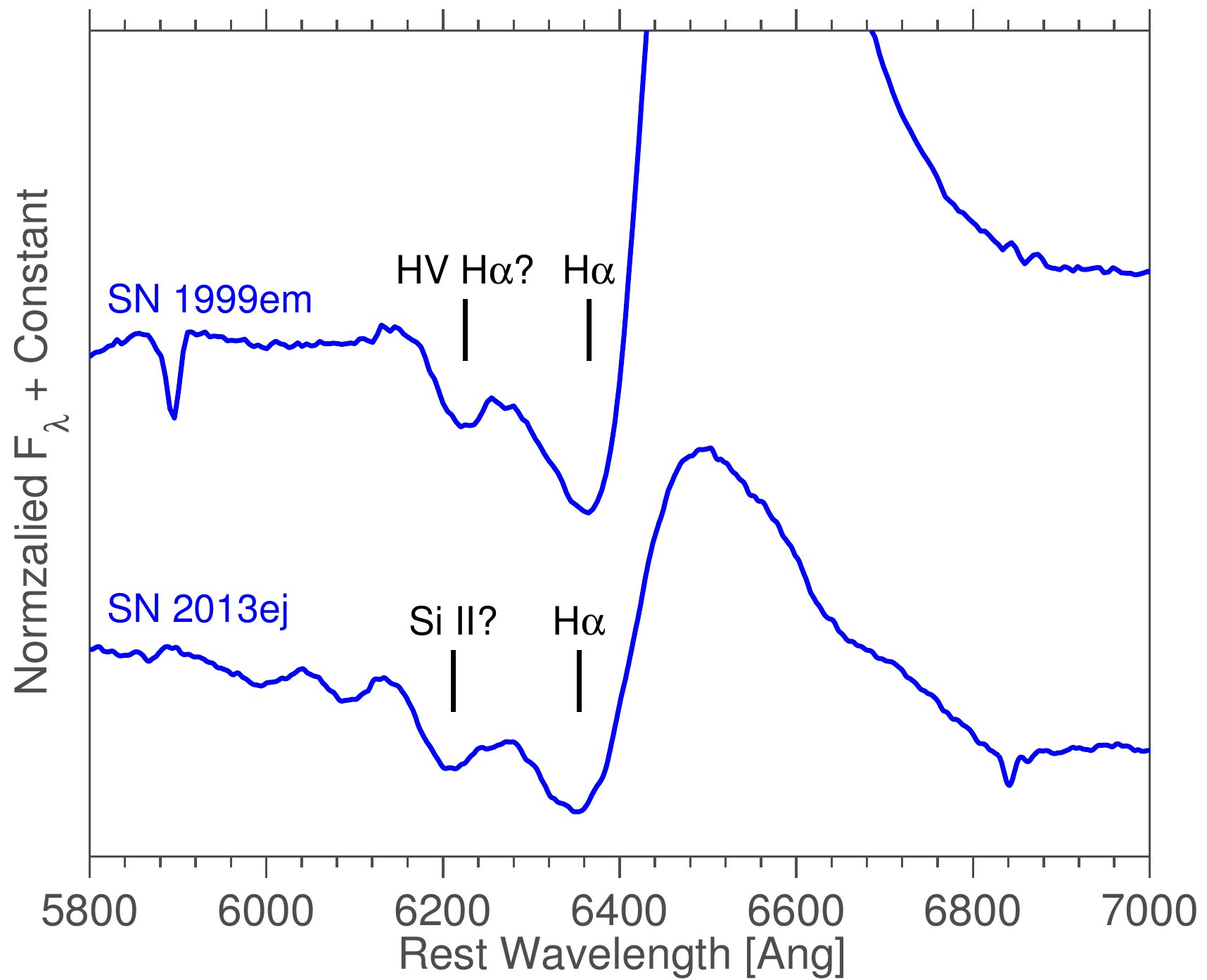}
\caption{The H$\alpha$ region in continuum-subtracted spectra of the Type IIP SNe 1999em \cite{Leonard2002} and 2013ej \cite{Valenti2013} (red). The absorption feature at $\sim6200\textrm{\AA}$ could be related to high velocity hydrogen, a sign of possible CSM interaction (as interpreted by Chugai, Chevalier and Utrobin 2007 \cite{Chugai2007} for SN\,1999em), or to Si II (as interpreted by Valenti et al. 2013 \cite{Valenti2013} for SN\,2013ej). This feature is only seen in some Type IIP SNe and only at some epochs (see also \cite{Pastorello2006, Inserra2012, Inserra2013}). Its nature is not yet clear.}
\label{fig:hvha}
\end{figure}

\subsection{Months From Explosion: The Nebular Phase}

Once the ejecta expand enough, they become completely transparent. At this phase the spectra display no continuum flux and only emission lines (Fig. \ref{fig:specs_nebular}). Spectra at these phases are useful for constraining the mass of the different line-emitting elements by measuring the flux in each line. These measurements can then be used to constrain the mass of the progenitor. Specifically, the [O~I] $6300\textrm{\AA}$ and $6364\textrm{\AA}$, Mg~I] $4571\textrm{\AA}$ and Na~I~D $5889\textrm{\AA}$ and $5895\textrm{\AA}$ line strengths correlate with progenitor mass \cite{Jerkstrand2014, Jerkstrand2014a}. 

Nebular spectra also reveal the heavy elements synthesized in the explosion (e.g. \cite{Kotak2006}). Furthermore, asymmetries in certain line profiles can indicate asphericity of the explosion (which is sometimes obscured by the sphericity of the envelope during the photospheric phase). 

Nebular spectra are thus valuable tools for constraining both SN progenitor properties and SN explosion models. 

\begin{figure}
\sidecaption
\includegraphics[width=\textwidth]{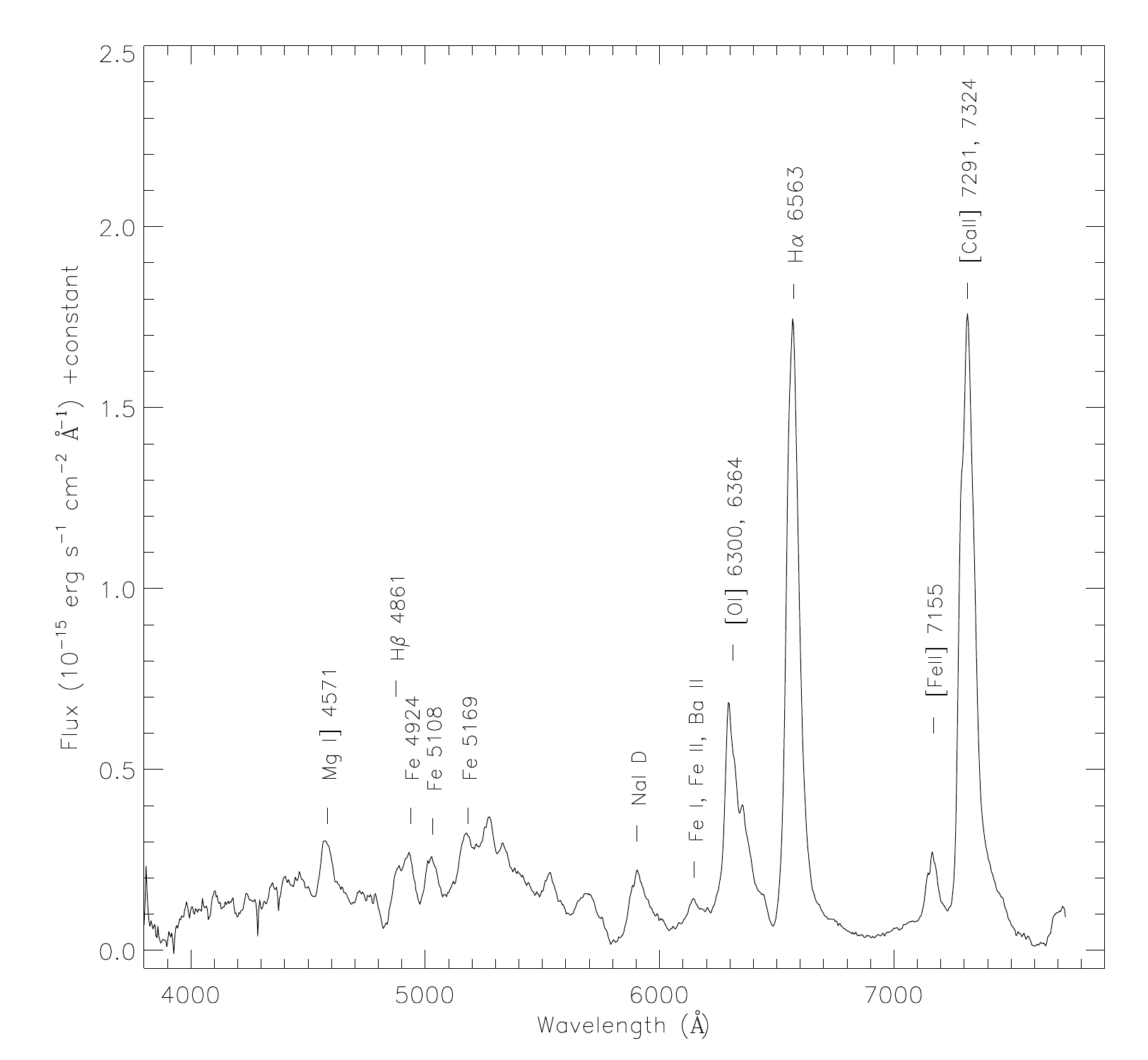}
\caption{A nebular spectrum of the Type IIP SN\,2004et, 401 days after explosion from \cite{Maguire2010}. Line strengths and profiles can be used to infer the progenitor mass and some otherwise-hidden explosion properties such as its nucleosynthetic yield and asymmetry.}
\label{fig:specs_nebular}
\end{figure}

\section{The Environments of H-rich SNe}
\label{sec:hosts}

Additional clues regarding the progenitor systems of the various SN types can be found by studying SN environments. 

One way of estimating the properties of a SN progenitor is to look at its neighboring stars and derive the star formation history at the SN site. The time elapsed since the peak of star formation can provide a rough estimate of the mass (longer time delays imply lower progenitor masses). The problem with this method is that it can be applied only to very nearby SN sites, where the stellar populations can be resolved. Badenes et al. (2009) \cite{Badenes2009} apply this method to the immediate environments of SN remnants in the Magellanic Clouds. However it is hard to determine the precise subtype of each explosion from the remnant alone.

The progenitor mass of SNe observed and typed directly, but at distances where their neighboring stellar populations can not be resolved, can be estimated by measuring the distance of the SN position to the nearest star-forming region (traced by blue/UV light or by H$\alpha$ emission), assuming that is the birth place of the progenitor. Larger distances imply longer lifetimes from birth to explosion and thus lower progenitor masses. While this method can not pinpoint the mass of any one progenitor with good accuracy, it can be used statistically to infer average differences in progenitor masses between different SN types.

Similarly, metallicity measurements of the nearest star-forming region can statistically connect progenitor composition to SN type. Direct metallicity measurements, however, are expensive as they require high signal to noise spectra of generally low surface brightness objects. Metallicity can be roughly estimated more easily by looking at the position of the SN in the galaxy. Most galaxies have metallicity gradients, with higher metallicities in the center of the galaxy and lower metallicities towards the outskirts (see \cite{Henry1999} for a review). An even more coarse metallicity estimate can be obtained by the total mass or luminosity of the galaxy. More massive (and hence more luminous) galaxies are generally more metal rich \cite{Tremonti2004}. Again, applied statistically to many SNe of different types, trends in SN position and global galaxy luminosity with SN type can be used to infer the role of metallicity in forming the different SN types.

Here I briefly summarize the progenitor constraints deduced from SN environment studies as they relate to H-rich SNe.

\subsection{Progenitor Mass}

All core collapse SNe are associated with starforming regions, indicating massive star progenitors. Badenes et al. (2009) \cite{Badenes2009} confirm this for core collapse SN remnants in the LMC.

For more distant galaxies, Type II events are found to be further away from their likely birth place compared to H-poor SNe \cite{Anderson2008, Galbany2014}. This indicates that H-rich SN progenitors generally have lower masses compared to H-poor SN progenitors. This result is consistent with direct detections of H-rich SN progenitors which are found to be in the mass range $10-17M_{\odot}$ (the lowest masses that produce core collapse SNe). Curiously, recent observations \cite{Anderson2012, Habergham2014} identify SNe IIn as the least associated with star forming regions of all H-rich events. This is in contradiction to indications of IIn SNe having very massive (several tens of solar mass) progenitors (e.g \cite{GalYam2007, Gal-Yam2009, Ofek2013, Ofek2014}). 

\subsection{Progenitor Metallicity}

Galaxy centers (where metallicity is high) are overabundant in H-poor SNe, while galactic (low-metal) outskirts preferentially host H-rich SNe \cite{VandenBergh1997, Tsvetkov2004, Anderson2009, Hakobyan2009}. This indicates that low metallicity could be important for H-rich SN progenitors to retain their hydrogen envelope prior to explosion. Direct metallicity measurements reach similar conclusions though with lower statistical significance \cite{Anderson2010}. 

Studies of global host-galaxy parameters (e.g. \cite{Prantzos2003, Prieto2008}) confirm the above picture by showing that H-poor SNe are less prevalent in faint, metal-poor host galaxies. Similar studies \cite{Arcavi2010, Arcavi2012} further show that Type IIb's prefer lower metallicity hosts compared to the general Type II SN population. Interestingly, Type Ib SNe (which are similar to Type IIb events post-peak) prefer high metallicity hosts (Fig. \ref{fig:host_fracs}). This suggests that both Type IIb and Ib progenitors could be otherwise-similar binary systems, with the differentiating factor being metallicity. In this scenario, the binary companion would be responsible for partially stripping the SN Ib/IIb progenitor, with metallicity doing the rest - at high metallicity, more hydrogen is stripped creating a Ib SN, whereas at low metallicity some hydrogen is retained in the envelope and a IIb SN ensues.

\begin{figure}
\sidecaption
\includegraphics[width=0.6\textwidth]{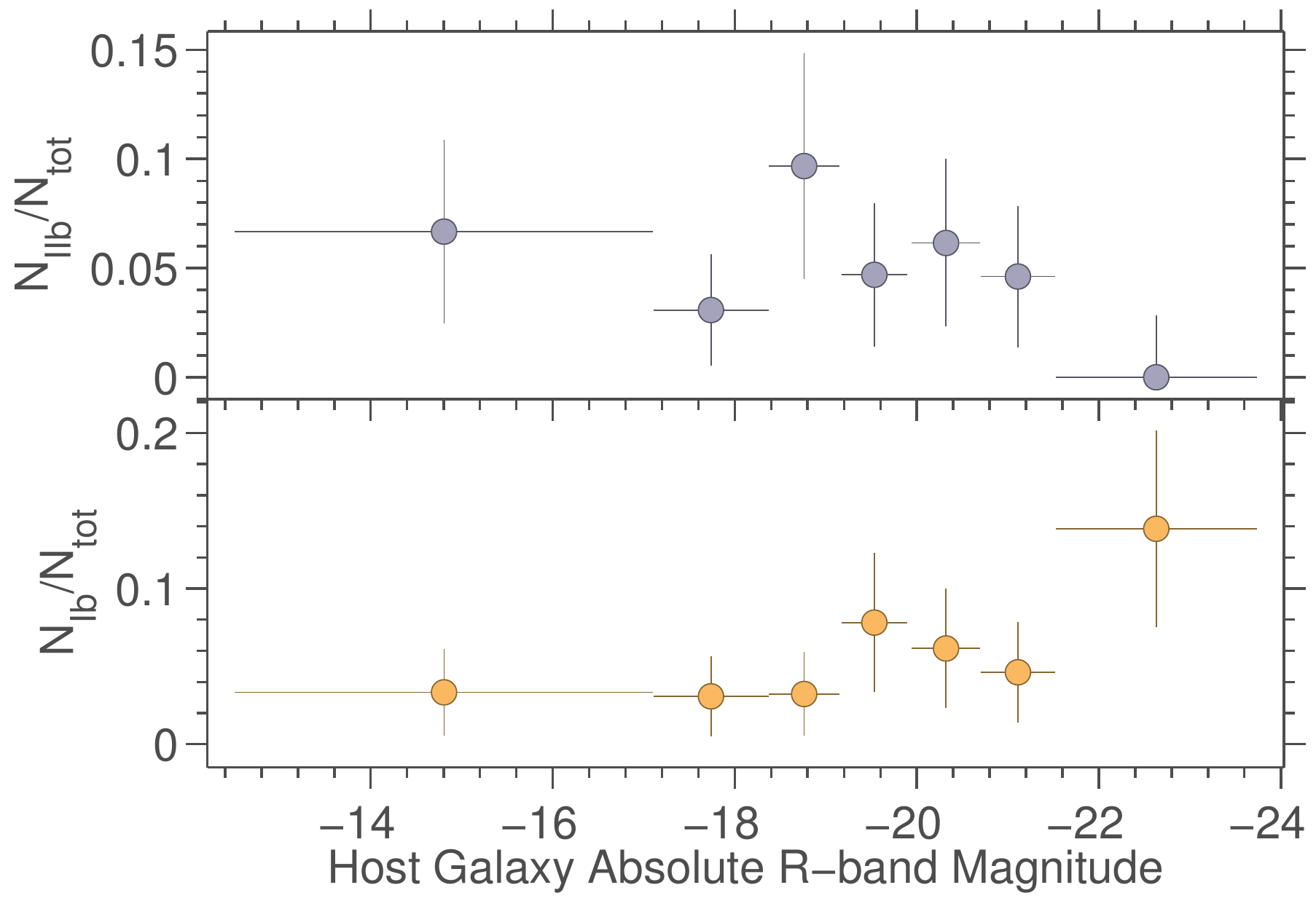}
\caption{Fractions of Type IIb SNe (top) and Type Ib SNe (bottom) relative to all core collapse SN types in different host-galaxy $R$-band magnitude bins. More luminous (and hence more metal-rich) galaxies host more SNe Ib and less SNe IIb. This indicates that metallicity may be the distinguishing factor between these two similar SN types. Figure adapted from \cite{Arcavi2012}.}
\label{fig:host_fracs}
\end{figure}

Finally, 87A-like SNe (including SN\,1987A itself) strongly prefer low (LMC-like) metallicity environments (e.g. \cite{Taddia2013}). This is consistent with the theoretical requirement for low metallicity to allow blue super-giants to explode \cite{Podsiadlowski1992}. 

The SN subtype relative rate dependence on galaxy luminosity has motivated many SN surveys to switch from searching for SNe in cataloged luminous galaxies, to performing wide-field untargeted searches that can find SNe also in uncatalogued dwarf galaxies.

\section{Open Questions}
\label{sec:qs}

Despite substantial advancement in our understanding of H-rich SNe since they were first identified, we still don't have a complete picture of their diversity and its mapping to progenitor channels. 

Some of the remaining open questions are listed below. Answering them could help us construct a more complete connection between the evolution of massive stars and their deaths as H-rich SNe.

\begin{itemize}
\item What powers the light curves of Type IIL SNe, and are their progenitors distinct from those of Type IIP SNe? Why do Type IIL SN spectra have less absorption in H$\alpha$?
\item Are all Type IIn SNe the explosions of very massive LBV stars? If so, why are they the least associated with star forming regions? 
\item How and when is the CSM in Type IIn SN progenitors formed?
\item Are all Type IIb SN progenitors binary systems? If not, how are their peculiar pre-explosion density structures formed? 
\item Why are Type IIP plateaus so uniform in length yet diverse in luminosity? 
\item What is driving the luminiosity-velocity correlation in Type IIP SNe?
\item What is the nature of the Si II / high-velocity H$\alpha$ feature observed in some Type IIP SNe?
\item How is the material seen in flash spectroscopy spectra formed and when was it ejected from the progenitor star?
\end{itemize}

New transient surveys and followup facilities now make it possible to collect larger samples of well-observed SNe than ever before. New robotic observing techniques are also allowing us to collect data on the first hours of a SN. Such datasets could answer these and other questions in the coming years.

\begin{acknowledgement}
I am grateful to N. Elias-Rosa, C. Glassman, D. Guevel, V. Hallefors, G. Hosseinzadeh, D. A. Howell, J. Jauregui, M. Modjaz, E. Nakar, F. Taddia and S. Valenti for providing information, materials, comments, and suggestions for the manuscript.
\end{acknowledgement}

\section{Cross-References}
\label{sec:xref}

\begin{itemize}
\item Supernova Progenitors Observed with HST
\item Interacting Supernovae
\item H-Poor Core Collapse Supernovae
\item Light Curves of Type II Supernovae
\item Progenitor of SN,1987A
\item Type Ia Supernovae
\item Shock Breakout Theory
\item Unusual Supernova and Alternative Power Sources
\item Superluminous Supernovae
\end{itemize}

\bibliographystyle{spphys}
\bibliography{refs}

%\section*{References}

\end{document}